\documentclass[twocolumns]{aa} % for a paper on 1 column    
\usepackage[varg]{txfonts}
\usepackage{amsmath}
\usepackage{graphicx}
\usepackage{siunitx}
\usepackage{comment}
\usepackage{xcolor}
\usepackage{multirow}
\usepackage{acronym}
\usepackage{xspace}
\usepackage{listings}
\usepackage{fontawesome}
\usepackage{subcaption}

\usepackage[colorlinks=true, allcolors=blue]{hyperref}

\usepackage{orcid}

\acrodef{GW}[GW]{gravitational wave}
\acrodef{BNS}[BNS]{binary neutron star}
\acrodef{H0}[$H_0$]{the Hubble constant}
\acrodef{EM}[EM]{electromagnetic}
\acrodef{CBC}[CBC]{compact binary coalescence}

\newcommand{\new}[1]{\textcolor{black}{#1}}

\newcommand{\icarogw}{\textsc{icarogw}\xspace}
\newcommand{\python}{\texttt{python} }

\newcommand{\jac}[1]{\frac{\mathrm{d}#1}{\mathrm{d}z}}

\newcommand{\de}{{\rm d}}
\usepackage{soul} % for strike through
\def\ie{{\emph{i.e.~}}}
\def\eg{{\emph{e.g.~}}}

\definecolor{codegreen}{rgb}{0,0.6,0}
\definecolor{codegray}{rgb}{0.5,0.5,0.5}
\definecolor{codepurple}{rgb}{0.58,0,0.82}
\definecolor{backcolour}{rgb}{0.95,0.95,0.92}

\lstdefinestyle{mystyle}{
    backgroundcolor=\color{backcolour},   
    commentstyle=\color{codegreen},
    keywordstyle=\color{magenta},
    numberstyle=\tiny\color{codegray},
    stringstyle=\color{codepurple},
    basicstyle=\ttfamily\footnotesize,
    breakatwhitespace=false,         
    breaklines=true,                 
    captionpos=b,                    
    keepspaces=true,                 
    numbers=left,                    
    numbersep=5pt,                  
    showspaces=false,                
    showstringspaces=false,
    showtabs=false,                  
    tabsize=2
}
\begin{document}

\title{\icarogw: A \python package for inference of astrophysical population properties of noisy, heterogeneous and incomplete observations}

\author{
Simone Mastrogiovanni\thanks{mastrosi@roma1.infn.it}$^{1}$ \orcidlink{0000-0003-1606-4183}, 
Grégoire Pierra$^2$ \orcidlink{0000-0003-3970-7970},
Stéphane Perriès$^2$ \orcidlink{0000-0003-2213-3579},
Danny Laghi$^3$ \orcidlink{0000-0001-7462-3794},
Giada Caneva Santoro$^4$ \orcidlink{0000-0002-0642-5507},
Archisman Ghosh$^5$ \orcidlink{0000-0003-0423-3533},
Rachel Gray$^{6,7}$ \orcidlink{0000-0002-5556-9873},
Christos Karathanasis$^{4}$ \orcidlink{0000-0002-0642-5507}
Konstantin Leyde$^8$ \orcidlink{0000-0002-0642-5507}
}

\institute{
$^1$ INFN, Sezione di Roma, I-00185 Roma, Italy\\ 
$^2$ Université Lyon, Université Claude Bernard Lyon 1, CNRS, IP2I Lyon/IN2P3,UMR 5822, F-69622 Villeurbanne, France \\
$^3$ Laboratoire des 2 Infinis - Toulouse (L2IT-IN2P3), Universit\'e de Toulouse, CNRS, UPS, F-31062 Toulouse Cedex 9, France\\
$^4$ Institut de Física d’Altes Energies (IFAE), Barcelona Institute of Science and Technology, Barcelona, Spain \\
$^5$ Department of Physics and Astronomy, Ghent University, Proeftuinstraat 86, 9000 Ghent, Belgium \\
$^6$ SUPA, University of Glasgow, Glasgow, G12 8QQ, United Kingdom \\
$^7$ Department of Physics \& Astronomy, Queen Mary University of London, Mile End Road, London, E1 4NS, United Kingdom \\
$^8$ Universit\'e Paris Cit\'e, CNRS, Astroparticule et Cosmologie, F-75013 Paris, France
}

\titlerunning{\icarogw}
\authorrunning{Mastrogiovanni et al.}

\date{\today}

\abstract{
We present \icarogw2.0, a pure CPU/GPU \python code developed to infer astrophysical and cosmological population properties of noisy, heterogeneous, and incomplete observations. \icarogw 2.0 is mainly developed for compact binary coalescence (CBC) population inference with gravitational wave (GW) observations. The code contains several models for masses, spins, and redshift of CBC distributions, and is able to infer population distributions as well as the cosmological parameters and possible general relativity deviations at cosmological scales. We present the theoretical and computational foundations of \icarogw 2.0, and we describe how the code can be employed for population and cosmological inference using \textit{(i)} only GWs, \textit{(ii)} GWs and galaxy surveys and \textit{(iii)} GWs with electromagnetic counterparts. We discuss the code performance on Graphical Processing Units (GPUs), finding a gain in computation time of about two orders of magnitudes when more than 100 GW events are involved for the analysis. We validate the code by re-analyzing GW population and cosmological studies, finding very good agreement with previous publications.}


\maketitle

%
%________________________________________________________________

\section{Introduction}
\label{sec:1}

\textbf{I}nferring \textbf{C}osmology and \textbf{A}st\textbf{R}ophysics with \textbf{O}bservations of \textbf{G}ravitational \textbf{W}aves (\href{https://github.com/simone-mastrogiovanni/icarogw}{\icarogw \faGithub}) is a pure \python code developed to infer the population properties of compact binary coalescences (CBCs) observed with gravitational waves (GWs). The problem of inferring the population properties from a sample of observations of astrophysical sources is a very common and long-standing problem shared among several research topics. With almost 100 GW observations from the last runs of the LIGO, Virgo, and KAGRA (LVK) collaboration~\citep{2021arXiv211103606T}, GW sources are moving rapidly to the ``population-inference'' domain. The first distribution of \icarogw was presented in the context of GW cosmology in \citet{2021PhRvD.104f2009M} and firstly used for the LVK analysis of \citet{gwtc3_H0} and released on the LVK official \href{https://git.ligo.org/cbc-cosmo/icarogw}{GitLab page \faGitlab}. The first version of \icarogw was also used in independent studies for population code validations \citep{Karathanasis:2022hrb,2023arXiv230212037T}, beyond General Relativity (GR) \citep{2022arXiv220200025L}, astrophysical processes \citep{2022arXiv220413495K}, and primordial black holes models~\citep{2023ApJ...943...29L,2023PhLB..83837720Z}. 

For astrophysics, population inference involves the correction of a \emph{selection bias}, or Malmquist bias \citep{1922MeLuF.100....1M}, that prevents the observation of a particular class of astrophysical processes, thus biasing the population analysis when not properly taken into account. Selection biases appear in many astrophysical observations involving neutrino physics \citep{Loredo:2001rx}, exoplanets \citep{2014ApJ...795...64F,2015ARA&A..53..409W}, galaxies and galaxy clusters \citep{Gonzalez_1997}, and $\gamma$-rays \citep{1998ApJ...502...75L}. For GW observations, selection biases are introduced by the sensitivity of the detector as a function of the GW frequency, which is also related to the binary parameters such as masses and redshift.

Besides the correction for the selection bias, GW population inference also has to account for the measurement of the source parameters, being uncertain -- since the GW signal is always observed in addition to noise in the data (data is \emph{heterogeneous}). 
Therefore, when reconstructing the population properties of GW signals, we face noisy, heterogeneous, and incomplete observations that require a specific statistical framework. \citet{2011AnApS...5.1657B} also refers to this type of analysis as \emph{extreme deconvolution}. 
Current techniques for population inference include the use of \emph{hierarchical Bayesian inference} \citep{2019MNRAS.486.1086M,2022hgwa.bookE..45V}. \icarogw provides a user-friendly \python environment to work with hierarchical Bayesian inference. As we will discuss later, only a few ``ingredients'' are required for population inference: \textit{(i)} a set of parameter estimation (PE) samples from the finite number of GW observations, \textit{(ii)} a set of simulations, or \textit{injections}, to calculate the detectable volume in parameter space and \textit{(iii)} a rate, or parameter population distribution.

\new{With respect to its first release, \icarogw 2.0 includes several useful improvements for all types of population studies. \icarogw 2.0 incorporates a user-friendly environment that allows the implementation of new population models without entering into the code details, and contains utilities to study the numerical stability of the Bayesian hierarchical inference.
For GW population studies, \icarogw 2.0 contains several cosmological and population models that are commonly used in previous literature. In App.~\ref{sec:cosmom} and App.~\ref{app:popmod}, we describe the CBC population models for spins, masses and redshift as well as cosmological expansion and modified gravity models. \icarogw 2.0 also contains three methodologies that can be used to perform GW cosmological studies, namely, (i) the spectral siren (ii) galaxy catalogue and (iii) the electromagnetic (EM) counterpart method.}

This paper is organized as follows.
In Sec.~\ref{sec:2} we summarise the theoretical and code implementation basics of hierarchical Bayesian inference. In Sec.~\ref{sec:3} we briefly discuss the infrastructure of the code and its \new{performance on GPUs}.
In Sec.~\ref{sec:4} we \new{validate the code by reproducing population and cosmological results published in literature from the third Gravitational Wave Transient Catalogue (GWTC-3) \citep{2021arXiv211103606T}}. In Sec.~\ref{sec:5} we draw our conclusions and discuss prospects for future development. 

\begin{figure*}
    \centering
    \includegraphics[scale=0.45]{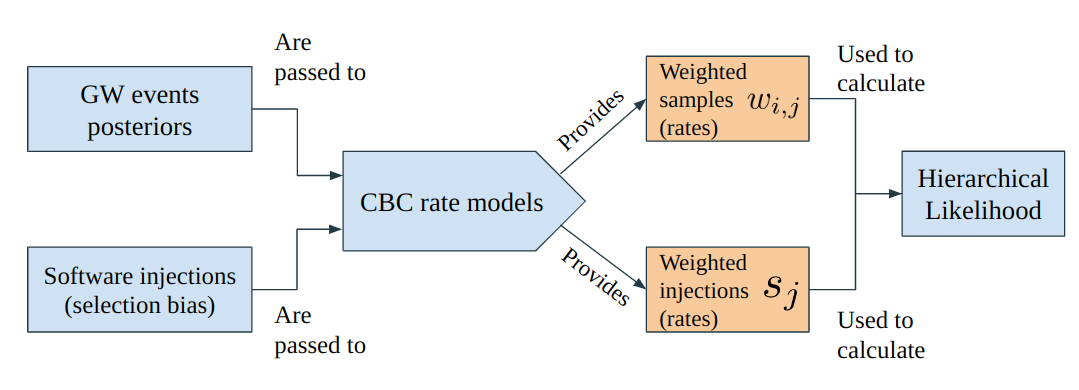}
    \caption{\icarogw logical structure. The orange boxes highlight the computation of the rate weights which are needed for the calculation of the hierarchical likelihood given in Eq.~\eqref{eq:hl_numer}.}
    \label{fig:icarogwflow}
\end{figure*}

\section{Inhomogeneous Poisson process and Bayesian inference}
\label{sec:2}

The main application of \icarogw is to infer the population parameters $\Lambda$ that describe the production rate of events in terms of \new{the GW} parameters $\theta$, namely $\frac{\de N}{\de t \de \theta}(\Lambda)$. For instance, in Sec.~\ref{sec:4} we will consider the rate of CBC in terms of source frame masses. The hierarchical likelihood of obtaining $N_{\rm obs}$ observations, each described by some parameters $\theta$, in a data collection $\{x\}$ for a given observing time $T_{\rm obs}$ from a population of events, with a constant rate and in presence of selection biases is given by (see \citet{2019MNRAS.486.1086M,2022hgwa.bookE..45V} for a detailed derivation):
\begin{equation}\label{eq:fund}
\begin{split}
    \mathcal{L}(\{x\}|\Lambda) &\propto e^{-N_{\rm exp}(\Lambda)} \prod_{i=1}^{N_{\rm obs}} \int  \mathcal{L}(x_i|\theta,\Lambda) \frac{\de N}{\de t \de \theta}(\Lambda) \de t \de \theta \\
    & \propto  e^{-N_{\rm exp}(\Lambda)} \prod_{i=1}^{N_{\rm obs}} T_{\rm obs} \int \mathcal{L}(x_i|\theta,\Lambda) \frac{\de N}{\de t \de\theta}(\Lambda) \de \theta.
\end{split}
\end{equation}
Eq.~\eqref{eq:fund} is often referred to as \textit{hierarchical likelihood}. By assuming a ``scale-free'' (i.e., neglecting information about the rate) prior $\pi(N_{\rm exp}) \propto 1/N_{\rm exp}$ on the expected number of detections, an equivalent form of Eq.~\eqref{eq:fund} can be derived:
\begin{equation}
    \mathcal{L}(x|\Lambda) \propto  \prod_{i=1}^{N_{\rm obs}} \frac{ \int \mathcal{L}(x_i|\theta,\Lambda) \frac{\de N}{\de t \de \theta} \de \theta}{ \int  p_{\rm det}(\theta,\Lambda) \frac{\de N}{\de t \de \theta} \de \theta}.
    \label{eq:fund_scalefree}
\end{equation}
Both the hierarchical likelihoods in Eq.~\eqref{eq:fund} and Eq.~\eqref{eq:fund_scalefree} contain several crucial quantities to the inference problem. In what follows, we will explain how \icarogw numerically computes these quantities and \new{obtains a numerical evaluation of} the full hierarchical likelihood defined in Eq.~\eqref{eq:fund}.

The first central quantity is the single-event likelihood $\mathcal{L}(x_i|\theta,\Lambda)$, \new{it quantifies the measurement uncertainties of the GW parameters $\theta$}. \new{The single-event likelihood is usually not directly accessible, instead}
we are provided with $N_{\rm s,i}$ \new{(PE)} posterior samples drawn from $p(\theta|x_i,\Lambda) \propto \mathcal{L}(x_i|\theta,\Lambda) \pi_{\rm PE}(\theta|\Lambda)$, where $\pi_{\rm PE}(\theta|\Lambda)$ is the prior used to generate the samples. \new{The PE samples represent the source parameters that we believe given the observation of data $x$.}
\icarogw numerically evaluates the likelihood integral in Eq.~\eqref{eq:fund} and in the numerator of Eq.~\eqref{eq:fund_scalefree} via Monte Carlo integration by summing over PE samples: 
\begin{equation}
    \begin{split}
     \int \mathcal{L}(x_i|\theta,\Lambda) & \frac{\de N}{\de t \de\theta}(\Lambda) \de \theta  \approx \\ & \frac{1}{N_{{\rm s},i}}  \sum_{j=1}^{N_{{\rm s},i}} \frac{1}{\pi_{\rm PE}(\theta_{i,j}|\Lambda)}\frac{dN}{dt d\theta}(\Lambda)\bigg|_{i,j} \equiv \frac{1}{N_{{\rm s},i}} \sum_{j=1}^{N_{{\rm s},i}} w_{i,j},
     \label{eq:intpe}
     \end{split}
\end{equation}
where the index $i$ refers to the event and the index $j$ to the posterior samples of the events. \new{The} weight $w_{i,j}$ is the number of CBC mergers happening per unit of time. As Eq.~\eqref{eq:intpe} is evaluated with a finite sum over posterior samples, we introduce the numerical stability estimator called \textit{effective number of posterior samples} per event $i$, as introduced in \citet{2023MNRAS.526.3495T}: 
\begin{equation}
    N_{{\rm eff},i}=\frac{(\sum_{j}^{N_{{\rm s},i}} w_{i,j}  )^2}{\sum_{j}^{N_{{\rm s},i}} w_{i,j}^2}.
    \label{eq:neffpe}
\end{equation}
This estimator quantifies how many samples per event are contributing to the evaluation of the integral. 
Typically, in population analyses such as \citet{gwtc3_H0}, it is required to have at least an effective number of posterior samples equal to 20 for each event and population model supported by the analysis. In case this requirement is not satisfied, \icarogw will artificially associate a null likelihood to the population model, as the model cannot be trusted.

The second central quantity is the \textit{expected number of events} $N_{\rm exp}(\Lambda)$, which is related to the selection bias and can be evaluated as: 
\begin{equation}
    N_{\rm exp}(\Lambda) = T_{\rm obs} \int p_{\rm det}(\theta,\Lambda) \frac{dN}{\de t \de \theta} \de \theta, 
    \label{eq:nexp}
\end{equation}
where $p_{\rm det}(\theta, \Lambda)$ is a detection probability that can be calculated as: 
\begin{equation}
    p_{\rm det}(\theta,\Lambda) = \int_{x \in \rm{detectable}} \mathcal{L}(x_i|\theta,\Lambda) \de x.  
\end{equation}
However, we do not have access to an analytical form of the detection probability, \new{unless some simplifying assumptions are made, see \citet{Gair:2022zsa} for an introductory example in the context of GW cosmology with galaxy catalogues}.
The current approach to evaluate selection biases is to use Monte Carlo simulations of injected and detected events \citep{LIGOScientific:2020kqk}, often shortly referred to as \textit{injections}. The injections are used to evaluate the \new{signal detectable} volume that can be explored in the parameter space and correct for selection biases. \new{The occurrence of detected injections}, proportional to $p_{\rm det}(\theta,\Lambda)$, and the population model used to generate them \new{can be used to evaluate the selection bias}.
\icarogw takes in input a set of $N_{\rm det}$ detected signals out of $N_{\rm gen}$ total injections generated from a prior $\pi_{\rm inj}(\theta)$ to calculate the integral in Eq.~\eqref{eq:nexp} using Monte Carlo integration:
\begin{equation}
    N_{\rm exp}  \approx \frac{T_{\rm obs}}{N_{\rm gen}} \sum_{j=1}^{N_{\rm det}} \frac{1}{\pi_{\rm inj}(\theta_j)}\frac{dN}{dt d\theta}\bigg|_j \equiv \frac{T_{\rm obs}}{N_{\rm gen}} \sum_{j=1}^{N_{\rm det}} s_j .
    \label{eq:nexpnum}
\end{equation}
Here we have again defined a weight $s_j$ with the dimension of a \new{CBC merger rate per detector time}. The injection prior $\pi_{\rm inj}(\theta)$ must be properly normalized to obtain a reasonable value of $N_{\rm exp}$, while a wrong normalization of $\pi_{\rm PE}(\theta)$ (which is used in Eq.~\eqref{eq:intpe}) will only result in an overall normalization factor to the overall hierarchical likelihood.
Following \citet{2019RNAAS...3...66F}, also for Eq.~\eqref{eq:nexpnum} we can define a numerical stability estimator, the \textit{effective number of injections}: 
\begin{equation}
    N_{\rm eff,inj} =  \frac{\left[\sum_j^{N_{\rm det}} s_j \right]^2}{\left[\sum_j^{N_{\rm det}} s_j^2 - 
 N_{\rm gen}^{-1} (\sum_j^{N_{\rm det}} s_j)^2 \right]}.
 \label{eq:neffinj}
\end{equation}
A typical value for numerical stability is $N_{\rm eff,inj}> 4N_{\rm obs}$.

In summary, the hierarchical likelihoods defined in Eq.~\eqref{eq:fund} and Eq.~\eqref{eq:fund_scalefree} are calculated using the approximations of the Monte Carlo integrals in Eq.~\eqref{eq:intpe} and Eq.~\ref{eq:neffinj}. The Monte Carlo integrated version of Eq.~\eqref{eq:fund} is
\begin{equation}
    \ln[\mathcal{L}(\{x\}|\Lambda)] \approx -\frac{T_{\rm obs}}{N_{\rm gen}} \sum_{j=1}^{N_{\rm det}} s_j + \sum_{i}^{N_{\rm obs}} \ln\left[ \frac{T_{\rm obs}}{N_{{\rm s},i}} \sum_{j=1}^{N_{{\rm s},i}} w_{i,j} \right]\,.
        \label{eq:hl_numer}
\end{equation}
while for the scale-free version, the Monte Carlo version is Eq.~\eqref{eq:fund_scalefree}
\begin{equation}
    \ln[\mathcal{L}(\{x\}|\Lambda)] \approx -N_{\rm obs} \ln\left[ \frac{1}{N_{\rm gen}} \sum_{j=1}^{N_{\rm det}} s_j\right] + \sum_{i}^{N_{\rm obs}} \ln\left[ \frac{1}{N_{{\rm s},i}} \sum_{j=1}^{N_{{\rm s},i}} w_{i,j} \right] \,.
        \label{eq:hl_numer_sf}.
\end{equation}

For each population model, \icarogw calculates one of the two likelihoods and the numerical stability estimators. If at least one of these numerical estimators is below the threshold set by the user, \icarogw returns a  $\ln[\mathcal{L}(\{x\}|\Lambda)]=-\infty $. This prevents the population model from being chosen.

\section{Code structure and GPU performance}
\label{sec:3}
\icarogw contains several \python modules used for population inference. In Fig.~\ref{fig:icarogwflow} we display a schematic view of \icarogw. 
PE samples and injections are stored in two different objects, \new{although they interact with common functions to calculate the CBC merger rate}. 
\new{The CBC rate models are the central objects of the code, they are \python classes that contain the production rate $\frac{\de N}{\de t \de \theta}$. Each specifies what are the event-level parameters $\theta$ used to calculate $\frac{\de N}{\de t \de \theta}$ and what are the population-level parameters $\Lambda$ used to calculate the rate model. The rate class also contains instructions on how to update the rate model from the population parameters $\Lambda$.}
\new{As we will describe more in details in Sec.~\ref{sec:4}, for GW cosmology, \icarogw distinguishes between detector frame frame parameters $\theta_{\rm D}$ and source frame parameters $\theta_{\rm S}$. Detector frame parameters are the CBC physical parameters observed on Earth, while the source frame ones are the actual parameters of the source.}
The GW parameters $\theta_{\rm D}$ that are necessary to calculate the CBC merger rate are passed to a CBC rate class that performs the conversion to source frame quantities $\theta_{\rm S}$ and calculates the weights $w_{i,j}$ and $s_j$ to be used for the calculation of the hierarchical likelihood and numerical stability estimators. Finally, the hierarchical likelihood computation is handled by a function embedded in the \python package \textsc{bilby} \citep{2019ApJS..241...27A,2019zndo...2602178A,2021MNRAS.507.2037A} for Bayesian inference. 

\begin{figure}[h!]
    \centering
    \includegraphics[scale=1]{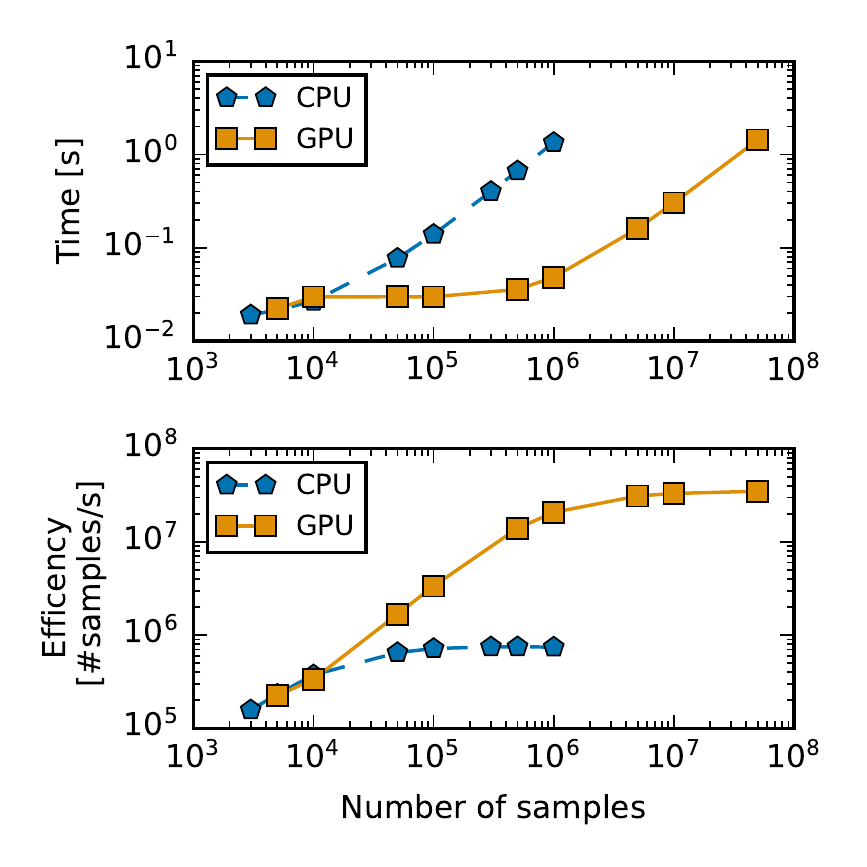}
    \caption{\textit{Top plot}: Average elapsed time taken by \icarogw to evaluate the hierarchical likelihood in Eq.~\ref{eq:fund}. \textit{Bottom plot:} Efficiency, defined as number of samples processed per second to evaluate the hierarchical likelihood in Eq.~\ref{eq:fund}. For both panels, the CPU is reported with \new{blue} pentagons and GPU with \new{orange} squares. The two panels are reported as a function of a number of samples, namely the sum of the number of PE samples and injections used for the computation. \new{The CPU markers are generated up to $10^6$ samples since at $10^5$ samples the efficiency is already saturated. Figures generated with double float precision on} CPU Intel Core i9-11950H (8 cores HT, 2.6-5.0GHz Turbo) and GPU NVIDIA GeForce RTX3080 (16Gb GDDR6, 6144 cores CUDA).}
    \label{fig:gpugain}
\end{figure}

Almost all the functions of \icarogw can also be executed on GPUs. The GPU is a specialized processor devoted to light and parallel computations. \icarogw uses \texttt{cupy} \citep{cupy_learningsys2017} to parallelize on GPU the computation of the weights $w_{i,j}$, $s_k$ and all the quantities that are computed from them. 

The use of the GPU greatly enhances the computation time needed for the calculation of the hierarchical likelihood. In Fig.~\ref{fig:gpugain} we show a comparison between the timing of the hierarchical likelihood in Eq.~\ref{eq:fund} computation with CPU and GPU using double float precision. The figure reports two figures of merit, the total time for the likelihood computation and the time taken per sample, in terms of the number of samples (PE+injections) used in the code. 
\new{The figures of merit have been generated on a computing machine employed to obtain the results of \citep{mastro2023} (see Fig.~\ref{fig:gpugain} for details). 
\newline
The likelihood's total time is a proxy for the time required to obtain a population fit as a function of the data size. From Fig.~\ref{fig:gpugain}, we can observe three regimes. For a number of samples $<10^4$, the CPU's and GPU's timing coincide, meaning that the computing time is not dominated by the sample size but by the internal \python routines. For a number of samples between $10^4$ and $10^6$, the CPU's time starts to be dominated by the sample size. In this middle regime, the GPU timing can be further improved by acting on routines that do not use data samples. The last regime is achieved for a number of samples higher than $10^6$, where both the CPU and GPU timings are dominated by the data samples. In agreement with \citep{2019PhRvD.100d3030T}, we find that the GPU enhance the computation time of a factor $\sim 100$ for population inference.
\newline
In \citep{mastro2023}, about $10^5$ samples were used on a CPU to obtain a population fit in 1 week of computation time. If $10^6$ samples on CPU are used, i.e. 10 times more GW events are considered, the total time required would be as high as 10 weeks of computation. Indeed 10 weeks of computation are not strictly computationally prohibitive but can certainly impact the number of models and exploratory studies that could be performed. 
\newline
The efficiency per sample is another way of quantifying when the CPU's and GPU's timings will start scaling with the number of samples. The efficiency per sample is defined as the data sample size divided by the total computing time.
Fig.~\ref{fig:gpugain} shows three regimes analogous to the total likelihood time. For a number of samples $<10^4$ the efficiency improves for both the CPU and GPU and the computing time is not dominated by the sample size. For a number of samples higher than  $10^4$, the CPU has already saturated its efficiency to a value of $10^{6}$ samples per second and the CPU's total time will start scaling linearly with the number of samples. Finally, for a total number of samples higher than $10^6$, the GPU's efficiency has saturated to a value of $10^{8}$ samples per second per sample and the total likelihood time will start scaling linearly as the samples pass.
\newline
Indeed we expect that different hardware configurations will display distinct values for the total likelihood time, efficiency, and for the transition between the three regimes. However, we expect the considerations discussed in the previous paragraphs to apply in general to all hardware facilities including GPU and CPU. Therefore, GPU computing will play a crucial role for future populations of GW events.
}

\section{Application to a compact binary coalescence case}
\label{sec:4}

Although \icarogw can be adapted to any problem involving an inhomogeneous Poisson process in the presence of selection bias, the code is mostly developed for GW science with CBCs. 
From GW detections we estimate the luminosity distance of the source $d_{\rm L}$, the detector masses $m_{1,d},m_{2,d}$, the sky position $\Omega$, and also parameters related to the spins of the compact objects (we indicate them generally with $\chi$ although two parameterizations are available, see App.~\ref{app:spinsmod}). \new{Therefore, the rate of CBC mergers as seen in detectors' data must be modelled as a function of $d_{\rm L}, m_{1,d},m_{2,d}$, and $\chi$, namely with a rate:}
\begin{equation}
    \frac{\de N}{\de d_{\rm L}  \de \Omega \de m_{1,d} \de m_{2,d} \de \chi \de t_{\rm d}},
    \label{eq:detrate}
\end{equation}
where $t_{\rm d}$ is the detector unit time. Eq.~\eqref{eq:detrate} is the CBC detector rate model. \new{The rate in Eq.~\ref{eq:detrate} does not strictly include any information about the Universe expansion model and its parameters. The information about cosmology can be included by rewriting the CBC merger rate in terms of source-frame parameters such as the source masses and redshift.}
\new{The source masses are related to the detector masses} through the relation 
\begin{equation}
m_{1/2,d}=m_{1/2,s}(1+z)    
\end{equation}
and the luminosity distance is related to the redshift by the choice of a cosmological model and possible GR deviations at cosmological scales. \icarogw contains several parameterizations for the luminosity distance that are reported explicitly with their ``population parameters'' in App.~\ref{app:cosmomod} for GR models and in App.~\ref{app:bgrmod} for modified gravity models.

\new{The CBC merger rate at the detector can be written as a function of source-frame parameters by performing a change of variables that involves the choice of the cosmological parameters, namely}
\begin{equation}
    \frac{\de N}{\de \theta_{\rm D} \de t_{\rm d}} = \frac{\de N}{\de \theta_{\rm S} \de t_{\rm s}} \frac{\de t_{\rm s}}{\de t_{\rm d}} \frac{1}{\det{J_{\rm D \rightarrow S}}}= \frac{\de N}{\de \theta_{\rm S} \de t_{\rm s}} \frac{1}{1+z} \frac{1}{\det{J_{\rm D \rightarrow S}}}.
    \label{eq:ratemodel}
\end{equation}
In the equation above, the factor $1/1+z$ comes from the difference between source-frame and detector-frame time, \new{and 
\begin{equation}
    \det{J_{\rm D \rightarrow S}} = \frac{\partial d_{\rm L}}{\partial z} (1+z)^2,
    \label{eq:specjac}
\end{equation}
is the Jacobian for the change of variables between the detector and the source frames. The expression of the differential of the luminosity distance can be found in App.~\ref{app:cosmomod} for standard cosmological models and in App.~\ref{app:bgrmod} for modified gravity models. The term
\begin{equation*}
    \frac{\de N}{\de \theta_{\rm S} \de t_{\rm s}}
\end{equation*}
is the CBC merger rate in the source frame and it is described by the distributions of CBCs in terms of redshift, spins and source frame masses. Several parameterizations of the CBC merger rate in the source frame are possible depending on the observational data available. We cover all these cases in the subsections below.}

\subsection{Spectral sirens merger rates}

The first case that we discuss is the ``spectral siren'' analysis \citep{Ezquiaga:2022zkx}. For this case, we are interested in inferring population properties of the source rate model, as well as cosmology and GR deviations from a population of GW events \emph{alone}. For this model, the detector event parameters $\theta_{\rm D}$ are $(d_{\rm L},m_{1,d},m_{2,d},\chi)$, \ie luminosity distance, detector masses, and spin parameters. The source event parameters are $\theta_{\rm S} = (z,m_{1,s},m_{2,s},\chi)$, \ie the redshift, two source masses, and the spin parameters. 
The detector rate for the spectral sirens analysis is parameterized as: 
\begin{equation}
    \begin{split}
    &\frac{\de N}{\de d_{\rm L}  \de m_{1,d} \de m_{2,d} \de \chi \de t_{\rm d}}= R_0 \Psi(z;\Lambda) \times \\ & p_{\rm pop}(m_{1,s},m_{2,s}|\Lambda)p_{\rm pop}(\chi|\Lambda) \frac{\de V_c}{\de z} \frac{1}{1+z} \frac{1}{\det{J_{\rm D \rightarrow S}}},
    \end{split}
    \label{eq:ratespecsiren}
\end{equation}
where $R_0$ is the CBC merger rate per comoving volume per year (in $\rm Gpc^{-3} yr^{-1}$), $\Psi(z;\Lambda)$ is a function parametrizing the rate evolution in redshift such that $\Psi(z=0;\Lambda)=1$, $p_{\rm pop}(m_{1,s},m_{2,s}|\Lambda)$ is a prior distribution describing the production of source masses, and $p_{\rm pop}(\chi|\Lambda)$ is a prior distribution for the production of spin parameters.  \icarogw implements the prior and rate distributions that are typically used for CBC population studies \new{and motivated by population syntheses studies of CBCs} \citep{gwtc3_H0}. App.~\ref{app:massmod}  lists \new{and describes} the models implemented for the masses, App.~\ref{app:spinsmod} the models for spins and App.~\ref{app:ratemod} the models for the merger rates as a function of redshift. Finally, $V_c$ is the comoving volume, \new{that also depends on the cosmological parameters.}

\subsection{Galaxy catalog merger rates}

The ``galaxy catalog'' \new{parametrization} adds information on the GW event redshift from galaxy surveys \citep{schutz,2012PhRvD..86d3011D, 2020PhRvD.101l2001G,Gray2022}. Also in this case, we are interested in inferring population properties of the source rate model, as well as cosmology and GR deviations from a population of GW events, this time with extra information coming from the galaxy catalog. 

In this method, the detector parameters $\theta_{\rm D} = (d_{\rm L},m_{1,d},m_{2,d},\Omega,\chi)$ are the luminosity distance and detector masses, sky direction pixel, and spins. The sky direction pixel area is measured in squared radians. The source event parameters are $\theta_{\rm S} = (z,m_{1,s},m_{2,s},\Omega,\chi)$, that is, the redshift and the two source masses, sky direction, and spins. \new{Here we directly provide the parametrization of the CBC merger rate for this method, for a more detailed derivation, see \citet{mastro2023}}. The detector rate for the galaxy catalog analysis is parameterized as:
\begin{eqnarray}
    \begin{split}
    &\frac{\de N}{\de d_{\rm L}  \de m_{1,d} \de m_{2,d} \de \Omega \de \chi \de t_{\rm d}} = R^{*}_{\rm gal,0} \frac{\Psi(z;\Lambda)}{1+z}\frac{p_{\rm pop}(m_{1,s},m_{2,s}|\Lambda)}{\det{J_{\rm D \rightarrow S}}} \\ & p_{\rm pop}(|\Lambda)   \bigg[ \frac{dV_c}{dz d\Omega} \phi_*(H_0)\Gamma_{\rm inc}(\alpha+ \epsilon+1,x_{\rm max}(M_{\rm thr}),x_{\rm min})  + \\ &  \sum_{j=1}^{N_{\rm gal}(\Omega)} \frac{f_{L}(M(m_j,z);\epsilon) p(z|z^j_{\rm obs},\sigma^j_{\rm z,obs})}{\Delta \Omega} \bigg].
    \label{eq:ratecat}
    \end{split}
\end{eqnarray}
where $R^{*}_{\rm gal,0}$ is the local CBC merger rate per galaxy per year (in $\rm yr^{-1}$). The sum of the two terms in the square brackets represents the galaxy number density in redshift and sky area that could host GW sources (see \citet{mastro2023} for more details). The first term is the \textit{completeness correction}, \ie it accounts for missing galaxies. It depends on the absolute magnitude threshold of galaxy detection $M_{\rm thr}$, on how likely more luminous galaxies can emit GW events (through the $\epsilon$ parameter), and on the Schecter luminosity function and its parameters $\phi_*$ and $\alpha$, $x_{\rm min/max}$ are defined in \citep{mastro2023} and are related to the minimum and maximum of the Schecter function. The second term in the square brackets is given by the galaxy distribution reported in the catalog. The function $f_{L}(M(m_j,z);\epsilon)$ quantifies how likely luminous galaxies emit GW events, while $p(z|z^j_{\rm obs},\sigma^j_{\rm z,obs})$ is the probability of having a certain value of $z$ given observed values of galaxy redshift inside the catalog.

\subsection{Sources with observed EM counterparts}

A third methodology to infer population and cosmological properties for GW events with an associated EM counterpart is also implemented. The model now takes into account additional constraints on sky position and redshift from a putative EM counterpart. \new{In this case, the CBC merger rate can still be parametrized as done in the previous sections, but the overall hierarchical likelihood should be modified to include the additional EM information.} We might be in two types of observational cases that we describe in the following.

\subsubsection{High-latency analyses}

In the case that we perform a ``high latency'' population analysis for the GW and EM observation, we expect GW studies to have produced PE samples on all the GW parameters such as sky position, redshift, and masses. In addition, the EM counterpart will provide us with an accurate sky localization of the source and a redshift localization with small uncertainties due to the peculiar velocities of the host galaxy.

By assuming that the GW measure is independent of the EM measure, the overall likelihood term is now $\mathcal{L}_{\rm EM+GW}(x_i|z,\Omega,m_{1,s},m_{2,s})$, which describes the measure of $z, \Omega, m_{1,s},m_{2,s}$ from EM and GW data.
Here we assume that the EM data measures $\Omega$ and $z$, while the GW data can measure $\Omega, z, m_{1,s},m_{2,s}$ independently, so that: 
\begin{equation}
\begin{split}
    \mathcal{L}_{\rm EM+GW}(x_i|&z,\Omega,m_{1,s},m_{2,s},\chi) \propto \\ &\mathcal{L}_{\rm EM}(x_i|z,\Omega) \mathcal{L}_{\rm GW}(x_i|z,\Omega,m_{1,s},m_{2,s},\chi).    
\end{split}
\end{equation}
The integral of the numerator in Eq.~\eqref{eq:fund_scalefree} becomes
\begin{equation}
    \begin{split}
    I = \int \mathcal{L}_{\rm EM}&(x_i|z,\Omega)  \mathcal{L}_{\rm GW}(x_i|z,\Omega,m_{1,s},m_{2,s},\chi) \\ & \frac{\de N}{\de z \de \Omega \de m_{1,s} \de m_{2,s} \de \chi \de t_{\rm s}} \frac{1}{1+z} \de m_{1,s},m_{2,s} \de \chi \de z \de \Omega.
    \end{split}
    \label{eq:Iz}
\end{equation}
Regarding the denominator of Eq.~\eqref{eq:fund_scalefree}, namely the selection bias, following the implementation of the same methodology in \citet{2020PhRvD.101l2001G}, we assume that the probability of detecting an EM counterpart (if present) given the GW detection is always 1. This assumption allows us to calculate the selection bias in Eq.~\eqref{eq:fund_scalefree} only using GW detectable signals, without the need to fold in a detection probability model for the EM counterpart. This assumption is physically valid only when the detection range of GW observatories is significantly lower than the detection range of EM observatories. As GW detectors improve their sensitivity this assumption should be revised as it could introduce a systematic bias on the estimation of $H_0$ \citep{2020PhRvL.125t1301C}.

\new{The integral in Eq.~\eqref{eq:Iz} can not be directly evaluated by calculating CBC rate weights $w_{i,j}$ and $s_lj$ as described in Sec.~\ref{sec:2}. Some additional steps need to be taken to account for the information from the EM counterpart.}
To perform the integral in Eq.~\eqref{eq:Iz}, \icarogw firstly identifies all the GW PE samples falling in the sky pixel in which we believe the EM counterpart is located. The PE selection procedure practically corresponds to performing the integral on $\de \Omega$ in Eq.~\eqref{eq:Iz}.
Next, a function of redshift $F(z)$ is defined from the selected GW PE samples. The function is
\begin{equation}
F(z) = \left[\frac{1}{N_s^{\rm EM}} \sum_{i}^{N_s^{\rm EM}} w_i\right] {\rm KDE}[z_i, {\rm weights}=w_i],
\end{equation}
where KDE is a kernel density estimate performed on the GW redshift samples $z_i$ with weights $w_i$. The weights are given by:
\begin{equation}
    w_i = \frac{1}{\pi_{\rm EM}(z^i) \pi_{\rm PE}(z^i,\vec{m}^i,\vec{\chi}^i)} \frac{\de N}{\de z \de m_{1,s} \de m_{2,s} \de \chi \de t_{\rm s}}\biggr|_{i} \frac{1}{1+z^i},
\end{equation}
where $\pi_{\rm EM}(z)$ is the prior used by the EM experiment to provide a redshift measure. The definition of $F(z)$, and the fact that we selected GW PE in the EM counterpart localization area, allows us to rewrite Eq.~\eqref{eq:Iz} as 
\begin{equation}
I \propto \int p_{\rm EM}(z|x_i,\Omega) F(z) \de z,
\end{equation}
where $p_{\rm EM}(z|x_i,\Omega)$ is EM posterior that localizes the EM counterpart in redshift given a certain sky area. The above integral can be evaluated with the usual Monte Carlo sum, by assuming a set of PE samples on $z$ drawn from the EM posterior as 
\begin{equation}
        I \approx \frac{1}{N_{\rm s , EM}}\sum_i^{N_{\rm s EM}} F(z_i).
\end{equation}

\begin{figure*}
    \centering
    \includegraphics[scale=1]{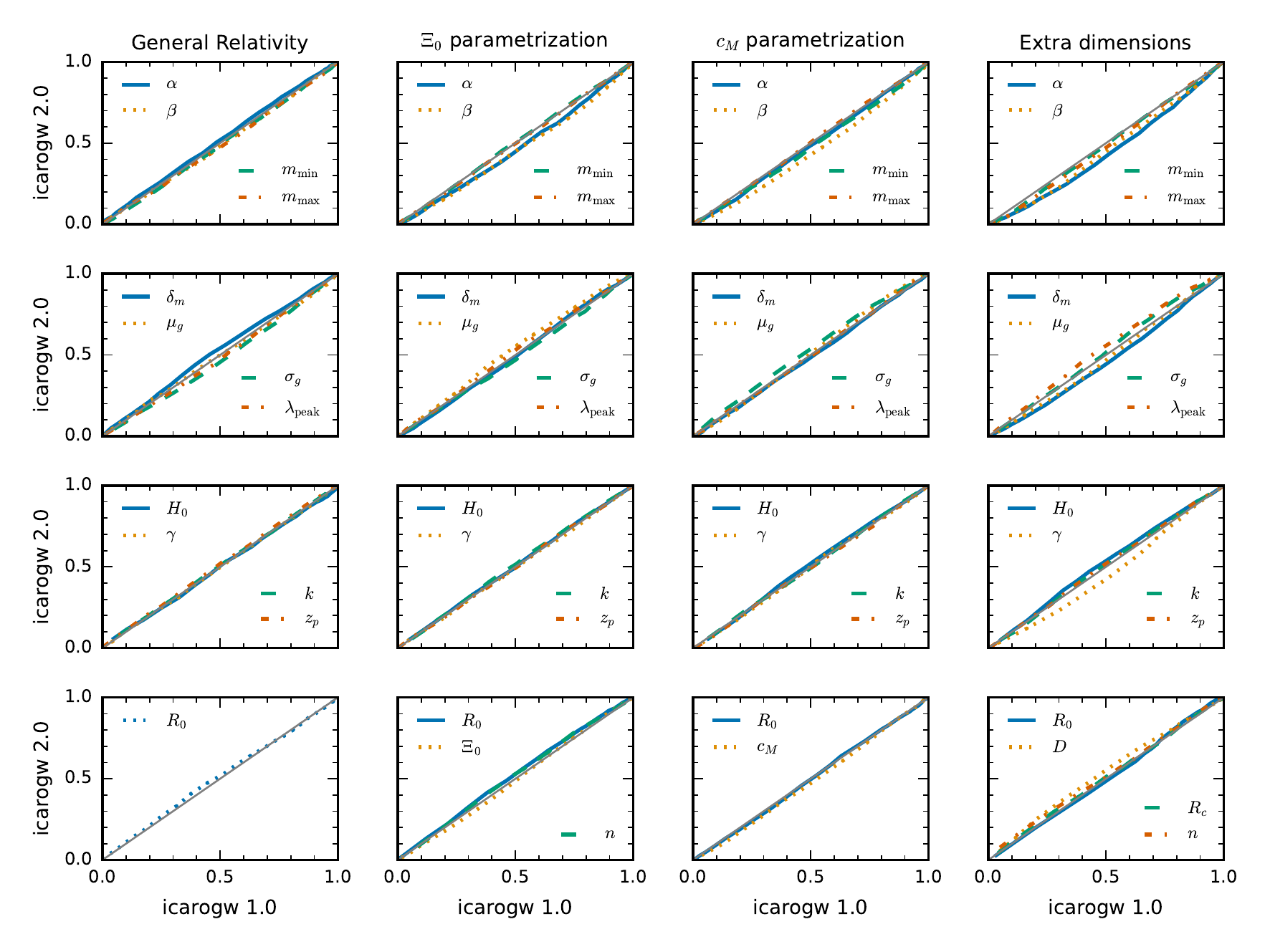}
    \caption{Quantile-Quantile plots of the marginal posterior distributions of the CBC merger rate population parameters inferred from the 42 BBHs in GWTC-3. The plots columns correspond to the model used to calculate the GW luminosity distance as a function of redshift, see App.~\ref{app:cosmomod}-\ref{app:bgrmod}. The legend and different lines report the population parameters. All the benchmark runs use the \textsc{power law + peak} model for the source mass (c.f. App.~\ref{app:massmod}) and the \textsc{madau-dickinson} merger rate model (c.f. App.~\ref{app:ratemod}).}
    \label{fig:spectralsiren_test_compact-label}
\end{figure*}

\subsubsection{Low-latency analyses}

In some cases, we might want to produce a ``low latency'' inference of population parameters (e.g. Hubble constant) given a GW and EM observation. In this case, we are typically provided with low-latency EM and GW low-latency skymap. The low-latency GW skymap can be thought of as a ``posterior'' $p_{\rm GW}(d_{\rm L},\Omega|x_i)$ of the luminosity distance and sky position \new{given the observed GW event}. The low-latency EM skymap can be thought of as a posterior $p_{\rm EM}(z,\Omega|x_i)$ for redshift and sky position. Also for this case, we make the assumption that the selection bias is dominated by the GW detection process and therefore no modelling for the EM selection bias is needed.

The GW-EM likelihood is a function only of sky position and redshift and can be written as
\begin{equation}
    \mathcal{L}_{\rm EM+GW}(x_i|z,\Omega) \propto \mathcal{L}_{\rm EM}(x_i|z,\Omega) \mathcal{L}_{\rm GW}(x_i|z,\Omega).    
\end{equation}
The integral in the numerator of the hierarchical Likelihood in Eq.~\eqref{eq:fund_scalefree} becomes
\begin{equation}
    I = \int \mathcal{L}_{\rm EM}(x_i|z,\Omega)  \mathcal{L}_{\rm GW}(x_i|z,\Omega)  \frac{\de N}{\de z \de \Omega \de t_{\rm s}} \frac{1}{1+z} \de z \de \Omega
\end{equation}
and the CBC merger rate is parameterized only in terms of redshift and sky position. We can now use the Bayes theorem to rewrite the above equation as
\begin{equation}
    I \propto \int \frac{p_{\rm EM}(z,\Omega|x_i)}{\pi_{\rm EM}(z,\Omega)}  \frac{p_{\rm GW}(d_{\rm L}(z,\Lambda),\Omega|x_i)}{\pi_{\rm GW}(d_{\rm L}(z,\Lambda),\Omega)}  \frac{\de N}{\de z \de \Omega \de t_{\rm s}} \frac{1}{1+z} \de z \de \Omega,
    \label{eq:25}
\end{equation}
where with $\Lambda$ we indicate any population parameter required to calculate the luminosity distance from redshift, while with $\pi_{\rm PE}$ and $\pi_{\rm GW}$ the priors used to generate the low-latency EM and GW skymaps. The integral now explicitly depends on the EM and GW low-latency skymaps. Eq.~\eqref{eq:25} can be evaluated analytically by drawing $N_{\rm EM}$ samples on $z$ and $\Omega$ the EM skymap, namely
\begin{equation}
    I \approx \frac{1}{N_{\rm EM}} \sum_{j=0}^{N_{\rm EM}} \frac{1}{\pi_{\rm EM}(z_j,\Omega_j)}  \frac{p_{\rm GW}(d_{\rm L}(z_j,\Lambda),\Omega_j|x_i)}{\pi_{\rm GW}(d_{\rm L}(z_j,\Lambda),\Omega_j)}  \frac{\de N}{\de z \de \Omega \de t_{\rm s}}\bigg|_j \frac{1}{1+z_j}.
    \label{eq:weiggg}
\end{equation}
The motivation for which the integral is evaluated by drawing samples from the EM skymap is that typically the EM source is better localized than the GW source. \new{By drawing samples from the EM skymaps, we have a higher probability for the weights in Eq.~\eqref{eq:weiggg} to be different from zero. Consequently, the numerical evaluation of $I$ is more stable.}

\section{Validation of the code}

We have performed several tests reproducing population results on CBC which are consistent with previous analyses. \new{In all the cases, we found that \icarogw 2.0 is able to reproduce results of previous literature.}

\subsection{Spectral siren analyses}

We used the same 42 BBHs used in \citet{gwtc3_H0} to reproduce populations, cosmological, and beyond-GR analyses generated with \icarogw 1.0. For the benchmarking, we employ a \textsc{powerlaw + peak} model (see App.~\ref{app:massmod}) for the source mass distribution of BBHs and a \textsc{madau-dickison} merger rate model (see App.~\ref{app:ratemod}). To evaluate selection biases, we used the procedure described in Sec.~\ref{sec:2} provided a set of detectable GW signals released with \citet{gwtc3_H0}.

The first analysis that we reproduced is the one presented in \citet{gwtc3_H0} which estimates conjointly population parameters for the BBH distribution together with the cosmological parameters $H_0, \Omega_m$ and $w_0$ (see App.~\ref{app:cosmomod} for more details). The first column of Fig.~\ref{fig:spectralsiren_test_compact-label} shows the quantile-quantile plots (QQ-plots) for the marginal posteriors on the population parameters obtained with \icarogw 1.0 and \icarogw 2.0. \new{When the two posterior are in agreement, the QQ-plot displays a bisector.}
For all the population parameters, we obtained an excellent agreement between the two posteriors \new{as shown by the first column of Fig.~\ref{fig:spectralsiren_test_compact-label}.}.

We also reproduced the constraints on modified gravity models and the population of BBHs generated with \icarogw 1.0 in \citet{2022arXiv220200025L}. The second, third, and fourth columns of Fig.~\ref{fig:spectralsiren_test_compact-label} show the QQ-plots for three modified gravity models (see Sec.~\ref{app:bgrmod} for more details). Also in this case the posteriors generated with \icarogw1.0 and \icarogw2.0 are in a good match between each other.

The tests with spectral sirens are also in good agreement with results obtained from the independent code \texttt{MGcosmoPop} \citep{michele_mancarella_2022_6323173} in \citet{Mancarella:2021ecn} (for the same set of BBH events). They are also in agreement with the results of \citet{2021PhLB..82236665E} using events from O3a modified gravity propagation and with the population-only results generated with the code \texttt{gwpopulation} \citep{2019PhRvD.100d3030T}.

\subsection{Galaxy catalog analysis} 

We tested \icarogw against the results generated by \texttt{gwcosmo} \citep{2020PhRvD.101l2001G,Gray2022} in \citet{gwtc3_H0} using the \textsc{glade+} \citep{Dalya:2021ewn} galaxy catalog with the infrared K-band. 

The procedure to construct a redshift distribution of possible host galaxies from the catalog is described in detail in \citet{mastro2023}. In summary, all the galaxies reported in \textsc{glade+} are divided into equal-sized sky pixels large 53 $\rm{deg}^2$. For each pixel, an apparent magnitude threshold is defined as the median of the apparent magnitude reported for all the galaxies. Galaxies fainter than the apparent magnitude threshold are removed from the catalog. Then, galaxies are subdivided into smaller pixels of 13 $\rm{deg}^2$ that are the ones used to run the analysis. The completeness correction is applied following the procedure described in \citet{mastro2023} that consists of assuming a Schecter function model for the galaxy distribution and counting all the galaxies missing due to the apparent magnitude threshold cut. The Schechter function model assumed for the K-band galaxies is specified by the parameters $M_{\rm min}=-27.85$, $M_{\rm max}=-19.84$, $\alpha=-1.09$,  $\phi_*=0.03 \,{\rm Mpc^{-3}}$ (for $H_0= 67.7$ [km/s/Mpc]).
For this benchmark analysis, following \citet{gwtc3_H0}, we assume that for each galaxy reported in the catalog the probability of hosting a GW event is proportional to the galaxy's intrinsic luminosity. The source mass and redshift population model is fixed to the same used by \texttt{gwcosmo} in \citet{gwtc3_H0}. 

\new{We only considered the inference for the Hubble constant $H_0$ in order to make a direct comparison with \citet{gwtc3_H0}.}
\begin{figure*}
    \centering
    \includegraphics[scale=1.1]{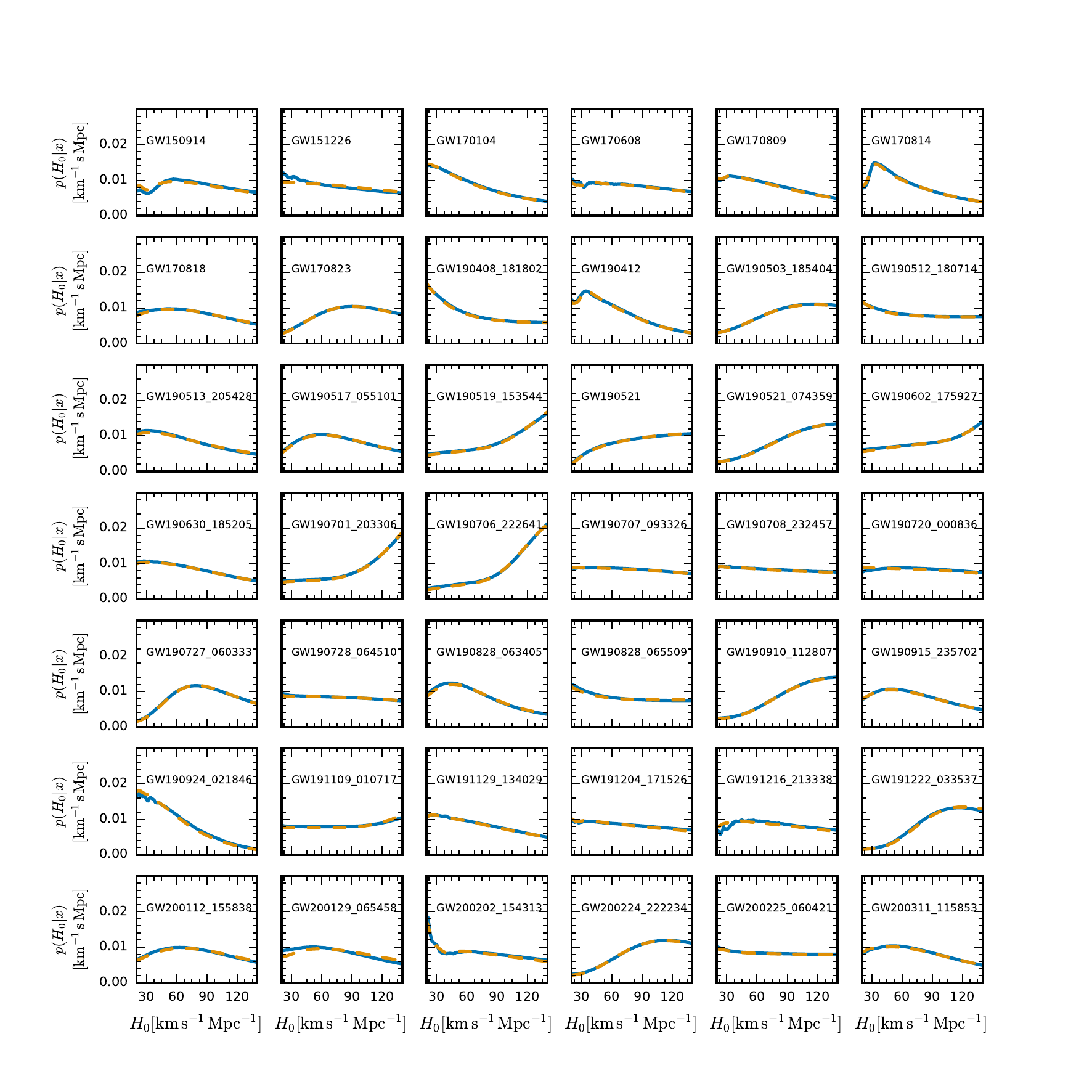}
    \caption{Posterior probability density distributions obtained by \icarogw2.0 (blue line) from 42 BBHs used in \citet{gwtc3_H0} in comparison with \texttt{gwcosmo} (orange dashed line).}
    \label{fig:O3_BBH_overall_cat}
\end{figure*}
The results that we obtained are shown in Fig.~\ref{fig:O3_BBH_overall_cat}. We find a good agreement for almost all the 42 BBHs present in the dataset. This is not surprising since most of the results are dominated by the assumption of the source mass distribution \citep{gwtc3_H0} \new{and as we discussed in the previous section, \icarogw 2.0 agrees well with previous literature when using the spectral siren method}.  However, note that the most close-by events such as GW150914, GW170814, and GW190824\_021846 are only partially dominated by population assumption on masses, and even for these events we obtain posteriors that are in agreement. 

We note that the posteriors we obtain for the catalog analysis are also consistent with the ones generated in \citet{2021JCAP...08..026F} from the code \href{https://github.com/CosmoStatGW/DarkSirensStat}{\texttt{
DarkSirensStat}}. However, the results in \citet{2021JCAP...08..026F} are generated with a different choice of BBHs and galaxy catalogue descriptions \new{compared to} the analyses performed in \citet{gwtc3_H0}.

We have also tested the $H_0$-catalog analysis on GW190814, one of the best localized and close-by dark sirens. For this event, we compare with the new version of \textsc{gwcosmo} \citep{2023arXiv230802281G}. We performed two analyses, the first one using only the GW event and the spectral siren method, the second using galaxies reported in the K-band from \textsc{glade+}. Differently from the previous analysis using the galaxy catalog, for GW190814 we use a pixel size (after computing the apparent magnitude cut) of 3 deg$^2$ because GW190814 localization area is a few tens of square degrees. Following \citep{gwtc3_H0}, we assume that GW190814 is an NSBH and use a source mass \textsc{Power law + peak} source mass model with the same population parameters used in \citep{gwtc3_H0}
\begin{figure}
    \centering
    \includegraphics[scale=1.0]{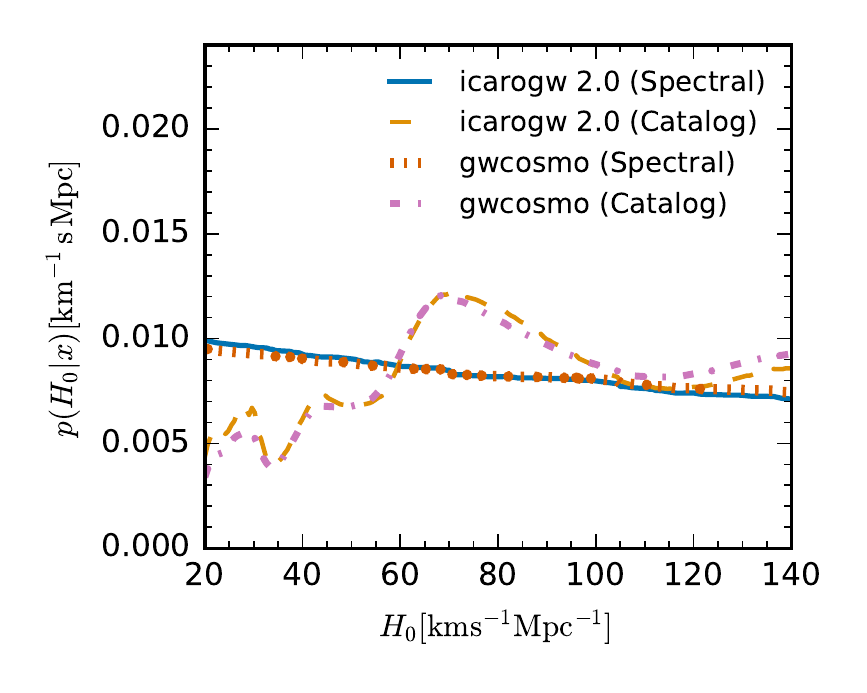}
    \caption{Hubble constant posteriors generated for the well-localized event GW190814. The plot reports results generated with \icarogw and the \textsc{gwcosmo} code for two cases: the spectral siren and galaxy catalog cases (see legend).}
    \label{fig:GW190814_comparison}
\end{figure}
Fig.~\ref{fig:GW190814_comparison} shows the comparison between \textsc{gwcosmo} and \textsc{icarogw} for the spectral and catalog analyses. The two results are in excellent agreement thus indicating the robustness of our analysis.

\subsection{Sources with observed EM counterparts} 

To test the electromagnetic counterpart method, we infer the Hubble constant using the BNS merger GW170817 and its EM counterpart. We perform two types of analyses, one using the ``high-latency'' approach and the other using the ``low-latency'' approach. For both cases, the injection set used to evaluate the selection biases is given by the BNS injection set released for O3 sensitivity in \citep{abbott2023population}.

For the high-latency approach we use \texttt{IMRPhenom PE} samples from \citet{LIGOScientific:2020kqk} consistently with the analysis of \textsc{gwcosmo}, while for the low-latency approach we use the GW170817 skymap publicly released for GWTC-2\footnote{\url{https://dcc.ligo.org/LIGO-P1800381/public}}. For both analyses, we used a sky position for the EM counterpart at right ascension $197.6$ deg and declination of $-23.4$ deg. Moreover, we assume a recessional velocity in the Hubble flow of $v_H= 3017$ km/s with uncertainty $\sigma_v=166$ km/s \citet{LIGOScientific:2020kqk}. 

The analysis done by the code \texttt{gwcosmo}  in \citep{LIGOScientific:2019zcs} used slightly different \new{population and cosmological} models to what it is currently implemented in \icarogw. \new{Therefore, to make a fair comparison, we implemented in \icarogw the population and cosmological models used in \citep{LIGOScientific:2019zcs}.}
First, the luminosity distance was approximated using linear cosmology, namely
\begin{equation}
  d_{\rm L} = d_c = \frac{cz}{H_0}.   
\end{equation}
The assumption corresponds to the following relations for the jacobians between source and detector frame and the differential of the comoving volume:
\begin{eqnarray}
  \frac{\partial d_{\rm L}}{\partial z} &=& \frac{c}{H_0}, \\ 
  \frac{\partial z}{\partial d_{\rm L}} &=& \frac{H_0}{c}, \\
  \frac{\partial V_c}{\partial z} &=& 4 \pi \frac{c^3 z^2}{H_0^3}. 
\end{eqnarray}
Second, The CBC merger rate model for GW170817 used in the analysis was uniform in detector masses, namely
\begin{equation}
    p(m_{1,d},m_{2,d})=\frac{\Theta(m_{2,d}<m_{1,d})}{2(m_{\rm d, max}-m_{\rm d, min})^2},
\end{equation}
where the $\Theta$ function ensures that the detector's secondary mass is lighter than the primary one. The analysis also neglected the $1/1+z$ factor from the difference between source and detector frames.  The overall merger rate was:
\begin{equation}
\begin{split}
& \frac{\de N_{\rm CBC}}{\de t_{\rm d} \de d_{\rm L} \de m_1 \de m_2} = \frac{\de N_{\rm CBC}}{\de t_{\rm d} \de d_{\rm L}} p(m_{1,d},m_{2,d}) = \\ & = R_0 \frac{\partial V_c}{ \partial z} \frac{\partial z}{ \partial d_{\rm L}} p(m_{1,d},m_{2,d}) = R_0 4\pi \frac{c^2 z^2}{H_0^2} \frac{\Theta(m_{2,d}<m_{1,d})}{2(m_{\rm d, max}-m_{\rm d, min})^2}.
\label{eq:rrmo}
\end{split}
\end{equation}

We remark that the aforementioned assumptions on cosmology and rate model are not expected to provide a noticeable difference when calculating the weights $w_{i,j}$ for the GW170817 PE samples. This is because GW170817 is a very close-by GW event, and even for extreme values of $H_0$, it remains at low redshift where the linear cosmology approximation is enough. Moreover, assumptions about the masses for GW170817 are not expected to strongly bias the result in the presence of an EM counterpart, as shown in \citep{2021PhRvD.104f2009M}.
\emph{However}, both masses and cosmological assumptions are expected to have an impact on the calculation of the selection bias. With O3 sensitivities, BNSs are detected up to a luminosity distance of $\sim 300-400$ Mpc, where the linear cosmology approximation can fail (especially if a high value of $H_0$ is chosen). So, even when reproducing GW170817 it is important to consider the rate model assumed in \citet{abbott2021population}. Thus, we implemented the rate model in Eq.~\ref{eq:rrmo} to study this comparison.

\begin{figure}
    \centering
    \includegraphics[scale=1.0]{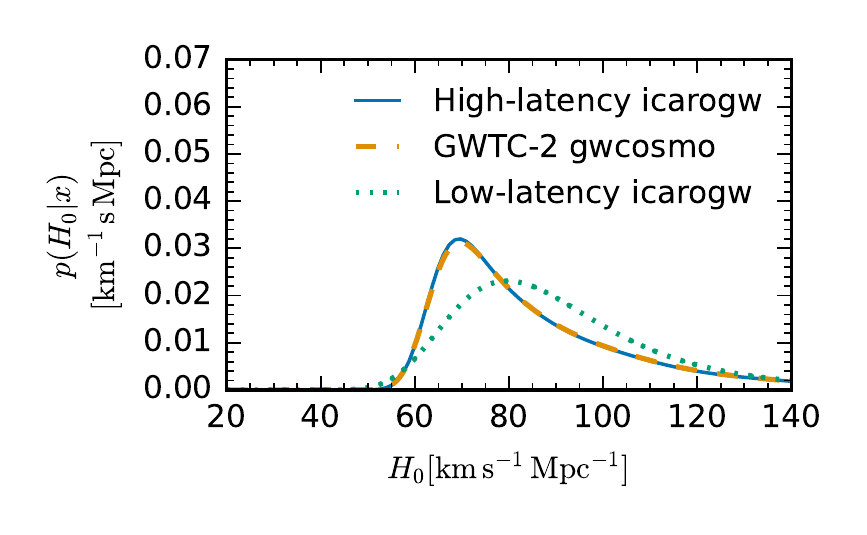}
    \caption{Posterior distributions for $H_0$ obtained from GW170817 with \icarogw2.0 with the high-latency approach (blue line) and low-latency approach (green dotted line) in comparison with \texttt{gwcosmo} using the high-latency approach (orange dash line). The posterior of \texttt{gwcosmo} is taken from \citet{LIGOScientific:2019zcs}.}
    \label{fig:EMcounterpart}
\end{figure}

\begin{figure}
    \centering
    \includegraphics[scale=1.0]{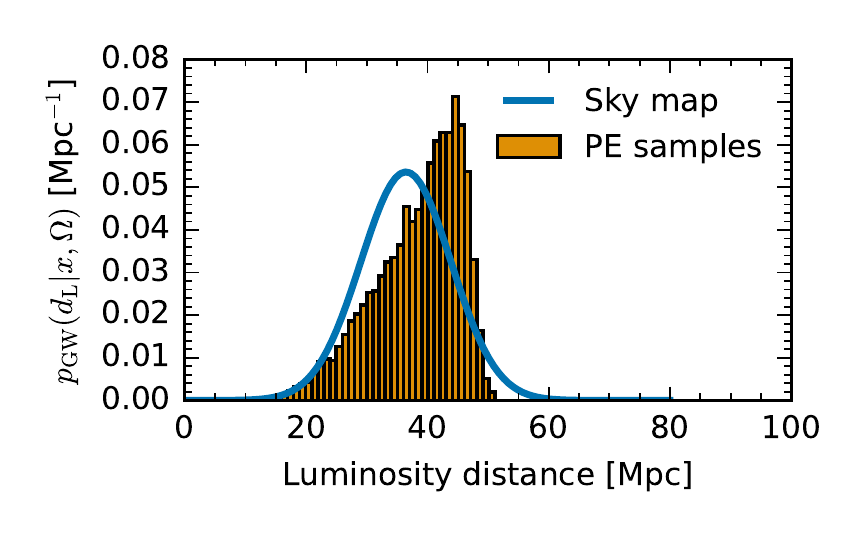}
    \caption{Luminosity distance posterior probability density function conditioned on the EM counterpart line-of-sight for GW 170817 using PE samples (orange histogram) and the low-latency skymap (blue line).}
    \label{fig:pos}
\end{figure}

Fig.~\ref{fig:EMcounterpart} shows the posteriors that we obtain with \icarogw2.0 and the method highlighted in Sec.~\ref{sec:4} for GW170817 in comparison with \texttt{gwcosmo}. The posteriors for the high-latency analyses are in good agreement with each other.
The main difference that we have found is that the low-latency posterior based on the GW skymap supports higher values of $H_0$. The result is motivated as follows. By fixing the sky localization of GW170817 EM counterpart, the GW PE samples and skymap have different support for the luminosity distance location. Fig.~\ref{fig:pos} shows that the skymap has a posterior distribution approximated as a Gaussian and it peaks at lower distance values if compared to the PE samples. At a fixed value of redshift for the EM counterpart, a lower luminosity distance can be fit using a higher value of $H_0$. This is the motivation for which the low-latency analysis supports slightly higher values of $H_0$. 
Concerning the high-latency analyses, the \icarogw analysis excellently agrees with the \textsc{gwcosmo} analysis of \citet{LIGOScientific:2019zcs}.

\section{Conclusions and future development}
\label{sec:5}

In this paper we have presented \icarogw2.0, a \python software for population properties inference in the presence of selection biases. We described some of the tests performed to check the validity of the CBC rate models implemented. We show that the results obtained by \icarogw 2.0 with the spectral sirens method are consistent with the previous version of the code (with and without the use of modified gravity models). We also showed that the galaxy catalog analysis matches results generated by the code \textsc{gwcosmo}. Finally, we showed how EM counterparts can be employed with the CBC rate models to infer the vale of the Hubble constant. We discussed two approaches from the EM counterpart case that produce consistent results with \textsc{gwcosmo}.

\icarogw2.0 can be easily adapted to any custom population inference problem involving the presence of noisy measurements and selection biases. Future development plans in GW science for \icarogw include more realistic models for CBCs that might include correlation among different variables (\eg mass and redshift), the inclusion of more beyond-GR models, and time-delay models.

The latest version of \icarogw2.0 is available to use in a public \href{https://github.com/simone-mastrogiovanni/icarogw}{GitHub \faGithub}  repository.

\section*{Software packages}

\icarogw uses the public \python packages \texttt{astropy} \citep{astropy:2022},
\texttt{bilby} \citep{2019ApJS..241...27A,2021MNRAS.507.2037A}, \texttt{cupy} \citep{cupy_learningsys2017}, \texttt{h5py} \citep{collette_python_hdf5_2014},   \texttt{healpy} \citep{2005ApJ...622..759G,Zonca2019},  \texttt{numpy} \citep{harris2020array}, \texttt{pickle} \citep{van1995python}, and \texttt{scipy} \citep{2020SciPy-NMeth} and their dependencies.

This paper has used plotting utilities from the \python packages \texttt{chainconsumer} \citep{Hinton2016} and \texttt{matplotlib} \citep{Hunter:2007}.

\begin{acknowledgements}
The authors are grateful for computational resources provided by the LIGO Laboratory and supported by National Science Foundation Grants PHY-0757058 and PHY-0823459.
This research has made use of data or software obtained from the Gravitational Wave Open Science Center (gwosc.org), a service of LIGO Laboratory, the LIGO Scientific Collaboration, the Virgo Collaboration, and KAGRA. LIGO Laboratory and Advanced LIGO are funded by the United States National Science Foundation (NSF) as well as the Science and Technology Facilities Council (STFC) of the United Kingdom, the Max-Planck-Society (MPS), and the State of Niedersachsen/Germany for support of the construction of Advanced LIGO and construction and operation of the GEO600 detector. Additional support for Advanced LIGO was provided by the Australian Research Council. Virgo is funded, through the European Gravitational Observatory (EGO), by the French Centre National de Recherche Scientifique (CNRS), the Italian Istituto Nazionale di Fisica Nucleare (INFN) and the Dutch Nikhef, with contributions by institutions from Belgium, Germany, Greece, Hungary, Ireland, Japan, Monaco, Poland, Portugal, Spain. KAGRA is supported by Ministry of Education, Culture, Sports, Science and Technology (MEXT), Japan Society for the Promotion of Science (JSPS) in Japan; National Research Foundation (NRF) and Ministry of Science and ICT (MSIT) in Korea; Academia Sinica (AS) and National Science and Technology Council (NSTC) in Taiwan. RG was supported by ERC starting grant SHADE 949572 and STFC grant ST/V005634/1.

\end{acknowledgements}

\appendix
\onecolumn

\section{Cosmological and GR deviation models}
\label{sec:cosmom}
Cosmological models and GR deviation models are handled by several classes and functions and are organized in high-level wrappers for quick use. They can be passed to the various  CBC rate models.

In Tab.~\ref{tab:cosmomod} we provide an overview of all the cosmological and GR models available. Note that GR deviation models are extensions of cosmological models, with beyond-GR population parameters on top of the cosmological background population parameters. GR deviation models only override the way in which the GW luminosity distance is computed (see next sections) while leaving the other cosmological quantities unchanged.

{\footnotesize	

\begin{longtable}{||c|c|c||}
\caption{Summary table for all the background cosmology and models available in \icarogw. More details on the models can be found in Sec.~\ref{app:cosmomod}.}
\label{tab:cosmomod}
\\
    \hline
    \multirow{2}{*}{\textbf{Model name}} & \multicolumn{2}{c||}{\textbf{Population parameters}} \\ \cline{2-3}
    & \textbf{Symbol} & \textbf{Description} \\ \hline
    
    \multirow{2}{*}{Flat $\Lambda$CDM} & $H_0$ & Hubble constant in [km/s/Mpc]  \\ \cline{2-3}
    &$\Omega_m$& Matter energy density \\ \hline
    
    \multirow{3}{*}{Flat $w$CDM} & $H_0$& Hubble constant in [km/s/Mpc]  \\ \cline{2-3}
    &$\Omega_m$& Matter energy density \\ \cline{2-3}
    &$w_0$& Dark Energy equation of state parameter \\ \hline
\end{longtable}
}

{\tiny

\begin{longtable}{||c|c|c||}
\caption{Summary table for all the background cosmology and beyond-GR models in \icarogw. More details on the models can be found in Sec.~\ref{app:bgrmod}.}
\label{tab:grmod}\\
    \hline
    \multirow{2}{*}{\textbf{Model name}} & \multicolumn{2}{c||}{\textbf{Population parameters}}  \\ \cline{2-3}
    & \textbf{Symbol} & \textbf{Description} \\ \hline
    
    \multirow{2}{*}{$\Xi_0$ model} & $\Xi_0$& See Eq.~\eqref{eq: def Xi parametrization dl}  \\ \cline{2-3}
    &$n$& See Eq.~\eqref{eq: def Xi parametrization dl} \\ \hline
    
    Running planck mass & $c_M$& See Eq.~\eqref{eq: def cm GW propagation for flat LCDM} \\ \hline
    
    \multirow{3}{*}{Extra dimensions} & $D$& \# spacetime dimensions, See Eq.~\eqref{eq:dpg}  \\ \cline{2-3}
    &$n$& Scaling parameter, See Eq.~\eqref{eq:dpg} \\ \cline{2-3}
    &$R_c$& GR deviation scale in Mpc \\ \hline
    
    \multirow{3}{*}{$\alpha$-log} & $\alpha_1$& See Eq.~\eqref{eq: def log parametrization dl}  \\ \cline{2-3}
    &$\alpha_2$& See Eq.~\eqref{eq: def log parametrization dl} \\ \cline{2-3}
    &$\alpha_3$& See Eq.~\eqref{eq: def log parametrization dl} \\ \hline
\end{longtable}

}

\subsection{Cosmological background models}
\label{app:cosmomod}

In principle \icarogw is able to use all the cosmologies included in \textsc{astropy}. However, for hierarchical inference, we have implemented only the models listed in the next subsections. For all the models we calculate the GW luminosity distance: 
\begin{equation}
    d_{\rm L}=\frac{c(1+z)}{H_0} \int_0^z \frac{\de z'}{E(z')},
\end{equation}
where $H(z)=H_0 E(z)$, which is the same as the EM luminosity distance assuming GR, as in this section. The differential of the luminosity distance is: 
\begin{equation}
    \frac{\partial d_{\rm L}}{\partial z}= \frac{d_{\rm L}(z)}{1+z}+\frac{c(1+z)}{H_0} \frac{1}{E(z)}.
\end{equation}
The comoving volume is:
\begin{equation}
    V_c=\int_0^z \de \Omega \de z' \frac{\de V_c}{\de z' \de \Omega},
\end{equation}
and the differential of the comoving volume is:
\begin{equation}
    \frac{\de V_c}{\de z \de \Omega}=\frac{c^3}{H^3_0} \frac{1}{E(z)} \left[\int_0^z \frac{\de z'}{E(z')} \right]^2.
\end{equation}
The function $E(z)$ depends on the cosmological model assumed. For the Flat $\Lambda$CDM model: 
\begin{equation}
E^2(z)=\Omega_{m}(1+z)^3+(1-\Omega_m),
\end{equation}
while for the  Flat $w_0$CDM model:
\begin{equation}
        E^{2}(z)=\Omega_{m}(1+z)^3+(1-\Omega_m) (1+z)^{3(1+w_0)}.
\end{equation}

\subsection{Beyond-GR models}
\label{app:bgrmod}

\begin{figure}
    \centering
    \includegraphics[scale=1.0]{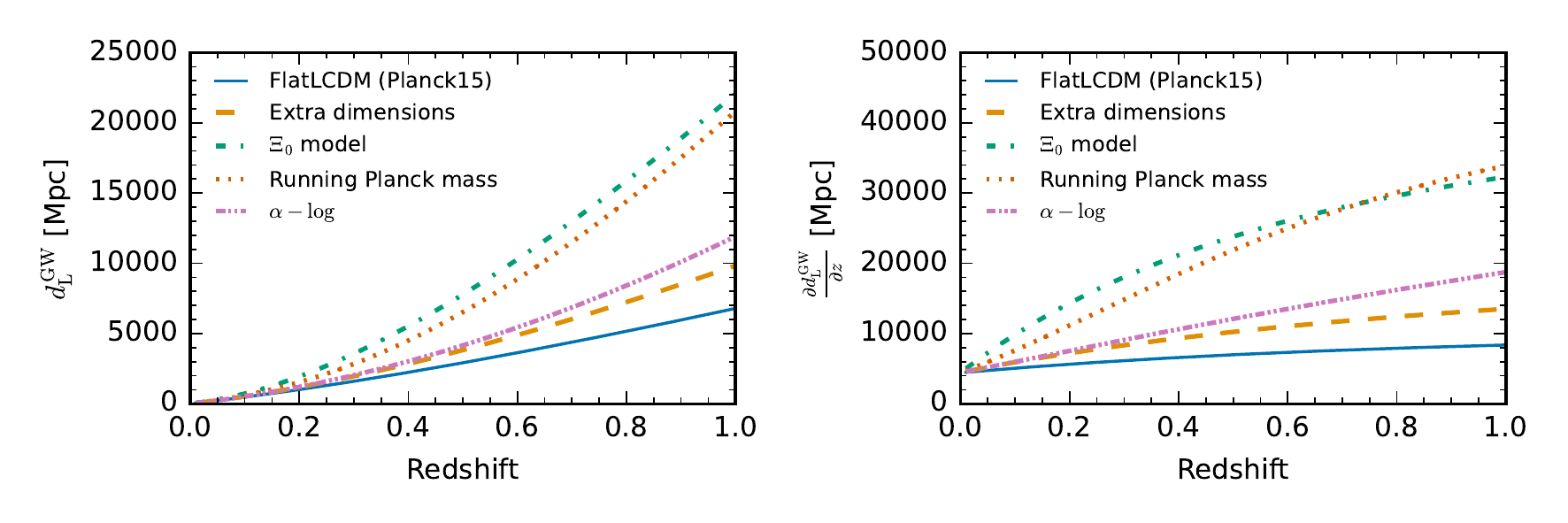}
    \caption{\textit{Left panel}: Luminosity distance as a function of redshift for the modified gravity models. \textit{Right panel:} Differential of the luminosity distance as a function of redshift for the modified gravity models. The lines indicate the models used to calculate the GW luminosity distance from the redshift.}
    \label{fig:modgrav}
\end{figure}

All the beyond-GR models implemented modify the luminosity distance, which we now refer to as $d_{L}^{\rm GW}$ (and its differential), while leaving untouched the comoving volume. We will refer to the standard luminosity distance as $d_{\rm L}^{\rm EM}$. In Fig.~\ref{fig:modgrav} we show how the luminosity distance and its differential with respect to redshift are modified for the models described below.

\subsubsection{The $\Xi_0$ model}

The luminosity distance is given by (see Eq.~2.31 of \cite{LISACosmologyWorkingGroup:2019mwx}):
\begin{equation}
    \label{eq: def Xi parametrization dl}
    d_{L}^{\mathrm{GW}} = d_{L}^{\mathrm{EM}}\left(\Xi_0 + \frac{1-\Xi_0}{(1+z)^n}\right)\,.
\end{equation}
The Jacobian is given by: 
\begin{equation}
    \label{eq: def Xi parametrization dl jacobian}
    \jac{d_{L}^{\mathrm{GW}}} = \jac{d_{L}^{\mathrm{EM}}}\left(\Xi_0 + \frac{1-\Xi_0}{(1+z)^n}\right) - d_{L}^{\mathrm{EM}}\frac{n(1-\Xi_0)}{(1+z)^{n+1}}\,.
\end{equation}

\subsubsection{The phenomenological $\log$ parametrization}
The luminosity distance is given by:
\begin{equation}
    \label{eq: def log parametrization dl}
    d_{L}^{\mathrm{GW}} = d_{L}^{\mathrm{EM}}\left[1+\sum_{\nu=1}^{n=3}\alpha_\nu\log^\nu(1+z)\right]\,.
\end{equation}
The Jacobian is given by: 
\begin{equation}
    \label{eq: def log parametrization dl jacobian ii}
    \jac{d_{L}^{\mathrm{GW}}} = \jac{ d_{L}^{\mathrm{EM}}}\frac{d_{L}^{\mathrm{GW}}}{d_{L}^{\mathrm{EM}}}
    +
    d_{L}^{\mathrm{EM}}\left[\sum_{\nu=1}^{n=3} \alpha_\nu \nu\frac{\log^{\nu-1}(1+z)}{1+z} \right]\,.
\end{equation}

\subsubsection{Extra-dimensions}
In the extra-dimensions model, the luminosity distance is given by (see Eq.~2.22 in \cite{Corman:2021avn}):
\begin{equation}
    d_{\rm L}^{\rm GW} = d_{\rm L}^{\rm EM}\left[1+\left(\frac{d_{\rm L}^{\rm EM}}{(1+z)R_c}\right)^n \right] ^{\frac{D-4}{2n}}\,.
    \label{eq:dpg}
\end{equation}
Let's define the following function:
\begin{equation}
    \mathcal{A}=\left[1+\left(\frac{d_{\rm L}^{\rm EM}}{(1+z)R_c}\right)^n \right]\,.
    \label{eq:dpg}
\end{equation}
We also define the exponential:
\begin{equation}
    \mathcal{E}=\frac{D-4}{2n}\,.
\end{equation}
We can write the previous equation as:
\begin{equation}
    \label{eq: def extra parametrization dl jacobian ii}
    \jac{d_{L}^{\mathrm{GW}}} = (\mathcal{A})^{\mathcal{E}}\left[\jac{ d_{L}^{\mathrm{EM}}}
    +
    \frac{n \mathcal{E}}{\mathcal{A}}
    \left(\frac{d_{\rm L}^{\rm EM}}{R_c}\right)^{n}
    \left(
    \jac{d_{L}^{\rm EM}}\frac{1}{(1+z)^n} -
    \frac{d_{\rm L}^{\rm EM}}{(1+z)^{1+n}}
    \right)\right]
    \,.
\end{equation}

\subsubsection{The $c_M$ parametrization}
Lastly, we consider a model with a running Planck mass \citep{Lagos:2019kds}: 
\begin{equation}
    d_{\rm L}^{\rm{GW}} = d_{\rm L}^{\rm{EM}} {\rm{exp}} \left[ \frac{c_M}{2} \int_0^{z} \frac{1}{(1+z')E^{2}(z')} dz' \right] \equiv d_{\rm L}^{\rm{EM}} {\rm{exp}}\left[ {\frac{c_M}{2} I(z)}\right],
\end{equation}
which defines $I(z)$.  In a flat $\Lambda$CDM model, $I(z)$ can be calculated analytically and the result is (Eq.~19 in \cite{Lagos:2019kds}):
\begin{equation}
    \label{eq: def cm GW propagation for flat LCDM}
    d_{\rm L}^{\rm GW} = d_{\rm L}^{\rm {EM}} {\rm{exp}} \left[\frac{c_M}{2\Omega_{\Lambda,0}} \ln \frac{1+z}{(\Omega_{\rm m,0}(1+z)^3+\Omega_{\Lambda,0})^{1/3}} \right]\,,
\end{equation}
otherwise it needs to be calculated numerically.
In any cosmology, the Jacobian is given by:
\begin{equation}
    \label{eq: def extra parametrization dl jacobian i}
    \jac{d_{L}^{\mathrm{GW}}} = \jac{ d_{L}^{\mathrm{EM}}}\frac{d_{L}^{\mathrm{GW}}}{d_{L}^{\mathrm{EM}}}
    +
    d_{L}^{\mathrm{GW}}\cdot
\frac{c_M}{2} \cdot
\frac{1}{(1+z) E^2(z)}.
\end{equation}

\section{CBC Population models}
\label{app:popmod}

Population models for the mass, redshift, and spins for CBC usually make use of probability density distributions. All the models currently available in \icarogw are not conditionally dependent on each other, \ie the probability distributions of redshift, source masses, and spins are independent from each other. The redshift, mass, and spin models provided can be used to construct a CBC rate model.

We provide a list of models for the CBC merger rate in redshift in Sec.~\ref{app:ratemod}, for source masses in  Sec.~\ref{app:massmod}, and for spins in Sec.~\ref{app:spinsmod}.

\begin{longtable}{||c|c|c||}
\caption{Summary table for all the merger rate models available in \icarogw. More details on the models can be found in Sec.~\ref{app:ratemod}.}
\label{tab:ratemod}
\\

    \hline
    \multirow{2}{*}{\textbf{Model name}} & \multicolumn{2}{c||}{\textbf{Population parameters}} \\ \cline{2-3}
    &  \textbf{Symbol} & \textbf{Description} \\ \hline
    
    Power Law rate & $\gamma$ & See Eq.~\eqref{eq:ratemod1}   \\ \hline

    \multirow{3}{*}{Madau Dickinson} &  $\gamma$ & See Eq.~\eqref{eq:ratemod2}  \\ \cline{2-3}
    &$k$& See Eq.~\eqref{eq:ratemod2} \\ \cline{2-3}
    &$z_p$& Redshift peak of merger rate \\ \hline
\end{longtable}

\begin{figure}
\centering
\begin{subfigure}{0.46\textwidth}
    \includegraphics[width=\textwidth]{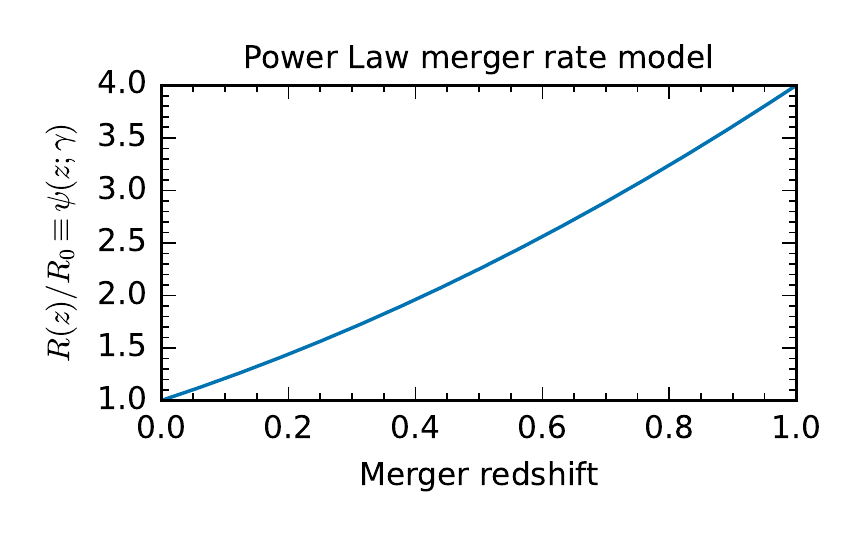}
    \caption{(a) Power Law rate model with $\gamma=2$.}
    \label{fig:first}
\end{subfigure}
\hfill
\begin{subfigure}{0.46\textwidth}
    \includegraphics[width=\textwidth]{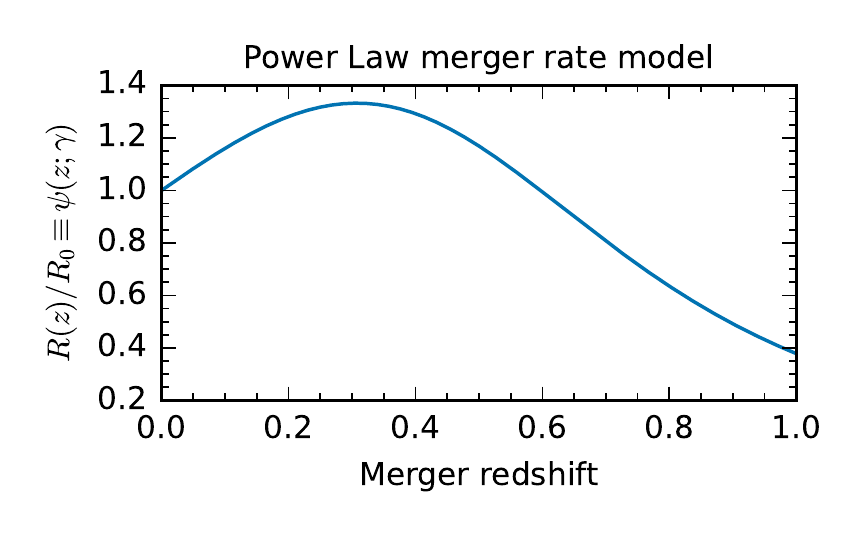}
    \caption{(b) Madau-Dickinson rate model with $\gamma=2.7, k=6, z_p=0.5$.}
    \label{fig:second}
\end{subfigure}
\caption{Sample of rate models implemented in \icarogw as a function of redshift.}
    \label{fig:ratemo}
\end{figure}

{\tiny

\begin{longtable}{||c|c|c||}
\caption{Summary table for all the background cosmology and models available in \icarogw.  The same mass models are also available for NSBH binaries. NSBH models also include three extra parameters: that are the minimum and maximum mass of the NS and the smoothing window for the lower end of the NS mass spectrum, respectively. More details on the models can be found in Sec.~\ref{app:massmod}.}
\label{tab:massmod}\\
    \hline
    \multirow{2}{*}{\textbf{Model name}} & \multicolumn{2}{c||}{\textbf{Population parameters}} \\ \cline{2-3}
    & \textbf{Symbol} & \textbf{Description} \\ \hline
    
    \multirow{4}{*}{Power Law} & $-\alpha$ &Power Law index primary mass  \\ \cline{2-3}
    &$\beta$& Power Law index secondary mass \\ \cline{2-3}
    &$m_{\rm min}$& Minimum source mass [$M_\odot$]\\ \cline{2-3}
    &$m_{\rm max}$& Maximum source mass [$M_\odot$]\\ \hline
    
    \multirow{8}{*}{Power Law + Peak} & $-\alpha$ &  Power Law index primary mass  \\ \cline{2-3}
    &$\beta$&Power Law index secondary mass \\ \cline{2-3}
    &$m_{\rm min}$& Minimum source mass [$M_\odot$]\\ \cline{2-3}
    &$m_{\rm max}$&Maximum source mass [$M_\odot$]\\ \cline{2-3}
    &$\delta_{\rm m}$& Smoothing parameter [$M_\odot$] Eq.~\eqref{eq:deltam}\\ \cline{2-3}
    &$\mu_g$& Peak of the gaussian  [$M_\odot$]\\ \cline{2-3}
    &$\sigma_g$& s.t.d of the gaussian  [$M_\odot$]\\ \cline{2-3}
    &$\lambda_{\rm peak}$& Fraction of events in gaussian $\in [0,1]$\\ \hline
    
    \multirow{7}{*}{Broken Power Law} & $-\alpha_1$ &  First Power Law index primary mass  \\ \cline{2-3}
    &$-\alpha_2$ & Second Power Law index primary mass  \\ \cline{2-3}
    &$\beta$& Power Law index secondary mass \\ \cline{2-3}
    &$m_{\rm min}$& Minimum source mass [$M_\odot$]\\ \cline{2-3}
    &$m_{\rm max}$& Maximum source mass [$M_\odot$]\\ \cline{2-3}
    &$\delta_{\rm m}$&  Smoothing parameter [$M_\odot$] Eq.~\eqref{eq:deltam}\\ \cline{2-3}
    &$b$& Defines $m_{\rm break}=b(m_{\rm max}-m_{\rm min})$\\ \hline
    
    \multirow{11}{*}{Multi Peak} & $-\alpha$ &  Power Law index primary mass \\ \cline{2-3}
    &$\beta$& Power Law index secondary mass \\ \cline{2-3}
    &$m_{\rm min}$&  Minimum source mass [$M_\odot$]\\ \cline{2-3}
    &$m_{\rm max}$& Maximum source mass [$M_\odot$]\\ \cline{2-3}
    &$\delta_{\rm m}$& Smoothing parameter [$M_\odot$] Eq.~\eqref{eq:deltam}\\ \cline{2-3}
    &$\mu_{g,\rm low}$& Peak lower mass gaussian [$M_\odot$]\\ \cline{2-3}
    &$\sigma_{g,\rm low}$& s.t.d. of the lower mass gaussian in [$M_\odot$]\\ \cline{2-3}
    &$\mu_{g,\rm high}$& Peak higher mass gaussian [$M_\odot$]\\ \cline{2-3}
    &$\sigma_{g,\rm high}$ & s.t.d higher mass gaussian in [$M_\odot$]\\ \cline{2-3}
    &$\lambda_{\rm g}$& Events in gaussians $\in [0,1]$\\ \cline{2-3}
    &$\lambda_{\rm g,low}$& Events in lower gaussian $\in [0,1]$\\ \hline
\end{longtable}
}

{\footnotesize
\begin{longtable}{||c|c|c||}
\caption{Summary table for all the merger spin models available in \icarogw. More details on the models can be found in Sec.~\ref{app:spinsmod}.}
\label{tab:spinmod}
\\
    \hline
    \multirow{2}{*}{\textbf{Model name}} & \multicolumn{2}{c||}{\textbf{Population parameters}} \\ \cline{2-3}
    & \textbf{Symbol} & \textbf{Description} \\ \hline
    
    \multirow{4}{*}{Default} & $\alpha_\chi$ & $\beta$-distribution parameter, see Eq.~\eqref{eq:B45}  \\ \cline{2-3}
    &$\beta_\chi$ & $\beta$-distribution parameter, see Eq.~\eqref{eq:B46}  \\ \cline{2-3}
    &$\sigma_t$ & Aligned spins dispersion, see Eq.~\eqref{eq:B47}  \\ \cline{2-3}
    &$\xi$ & Fraction of events with semi-aligned spins\\ \hline
    
    \multirow{5}{*}{gaussian}& $\mu_{\xi,{\rm eff}}$& Mean of $\chi_{\rm eff}$  \\ \cline{2-3}
    &$\mu_{\xi,{\rm p}}$& Mean of $\chi_{\rm p}$ \\ \cline{2-3}
    &$\sigma_{\xi,{\rm eff}}$& s.t.d of $\chi_{\rm eff}$\\ \cline{2-3}
    &$\sigma_{\xi,{\rm p}}$& s.t.d of $\chi_{\rm p}$\\ \cline{2-3}
    &$\rho$& $\chi_{\rm eff}-\chi_{\rm p}$ correlation term\\  \hline
\end{longtable}
}

\subsection{CBC redshift rate evolution models}
\label{app:ratemod}

\icarogw contains two models for the redshift evolution of the merger rate, see Eqs.~\eqref{eq:ratespecsiren}-\eqref{eq:ratecat}. Table \ref{tab:ratemod} summarises the merger rate redshift models, while Fig.~\ref{fig:ratemo} provides some examples of the models for specific values of the parameters.

\subsubsection{Power Law}

The rate is parametrized as: 

\begin{equation}
    \psi(z;\gamma)=(1+z)^\gamma\,.
    \label{eq:ratemod1}
\end{equation}

\subsubsection{Madau-Dickinson}

The rate is parametrized following the star formation rate of \citet{Madau:2014bja} as:

\begin{equation}
    \psi(z;\gamma)=[1+(1+z_p)^{-\gamma-k}] \frac{(1+z)^\gamma}{1+\left(\frac{1+z}{1+z_p}\right)^{\gamma+k}}\,.
    \label{eq:ratemod2}
\end{equation}

\subsection{Mass models}
\label{app:massmod}
Most of the mass models are composed of gaussian and power law distributions which we report in the following. The simple truncated power law distribution is given by

\begin{eqnarray}
\label{TPL}
 \mathcal{P}(x|a,b,\alpha)= \bigg\{
\begin{array}{lr}
     \frac{1}{N}x^{\alpha}, & \quad  \left(a<x<b\right) \\
     0, & \quad \left({\rm {otherwise}}\right) 
\end{array} 
\end{eqnarray}
and the  normalization factor is given by:
\begin{eqnarray}
N= \bigg\{
\begin{array}{lr}
  \dfrac{1}{\alpha+1}[b^{\alpha+1}-a^{\alpha+1}], & \quad {\rm{if \, \alpha \neq-1}}\\
    {\ln}{\dfrac{b}{a}}, & \quad  \rm{if \, \alpha =-1}
    \end{array}
\end{eqnarray}
The truncated gaussian distribution is given by:
\begin{eqnarray}\label{eqn:trGauss}
\mathcal{G}_{[a,b]}(x|\mu,\sigma)=\bigg\{
\begin{array}{lr}
\dfrac{1}{N}\dfrac{1}{\sigma\sqrt{2\pi}} e^{-\frac{(x-\mu)^2}{2\sigma^2}}, &\quad  a<x<b \\
0, &\quad \text{otherwise}
\end{array}
\end{eqnarray}
The normalization factor is expressed through the error function. %To see how, change variable in $t=\dfrac{x-\mu}{\sigma\sqrt{2}}$. 
Then the normalization factor is:
\begin{equation}
    N=\int_{a}^{b}\frac{1}{\sigma\sqrt{2\pi}} e^{-\frac{(x-\mu)^2}{2\sigma^2}}dx=\int_{(a-\mu)/(\sigma\sqrt{2})}^{(b-\mu)/(\sigma\sqrt{2})}\frac{1}{\sqrt{\pi}} e^{-t^2}dt.
\end{equation}
Using the symmetry of the integrand around $x=\mu$ ($t=0$) and the definition of erf function\footnote{ \href{https://docs.scipy.org/doc/scipy-0.14.0/reference/generated/scipy.special.erf.html}{\textsc{Scipy} function.}}:
\begin{equation}
    \mathrm{erf}(z)=\frac{2}{\sqrt{\pi}}\int_0^z e^{-t^2}dt,
\end{equation}
it follows that:
\begin{equation}
N = \frac{1}{2} \left( {\rm{erf}} \left[\frac{b-\mu}{\sigma \sqrt{2}} \right] - {\rm{erf}} \left[\frac{a-\mu}{\sigma \sqrt{2}} \right] \right).
\end{equation}

In \icarogw, we factorize the prior on mass as: 
\begin{equation}
\pi(m_{1,s},m_{2,s}|\Lambda)=\pi(m_{1,s}|\Lambda)\pi(m_{2,s}|m_{1,s},\Lambda)\,.
\label{massprior0}
\end{equation}
When dealing with a NSBH, the neutron star is assigned to $m_{2,s}$ and the distribution of $m_{2,s}$ will be a simple power law defined between a minimum and a maximum mass, which are different from the ones assumed for the black hole.

In some of the models, we also apply a smoothing factor to the {\it lower} end of the mass distribution at $m=m_{\rm min}$:
\begin{equation}
\pi(m_{1,s},m_{2,s}|\Lambda)=[\pi(m_{1,s}|\Lambda)\pi(m_{2,s}|m_{1,s},\Lambda)]S(m_1|m_{\rm min},\delta_m)S(m_2|m_{\rm min},\delta_m),
\label{massprior}
\end{equation}
where $S$ is a sigmoid-like window function (Eqs.~B6-B7 of \cite{Abbott:2020gyp}): 
\begin{eqnarray}
\label{eq:smoothing}
S(m_{1,s} | m_{\rm min}, \delta_m) = \bigg\{
\begin{array}{lr}
    0, & \left(m< m_{\rm min}\right) \\
    \left[f(m - m_{\rm min}, \delta_m) + 1\right]^{-1}, & \quad \left(m_{\rm min} \leq m < m_{\rm min}+\delta_m\right) \\
    1, & \quad \left(m\geq m_{\rm min} + \delta_m\right)
\end{array}
\end{eqnarray}
with
\begin{equation}
    f(m', \delta_m) = \exp \left(\frac{\delta_m}{m'} + \frac{\delta_m }{m' - \delta_m}\right).
    \label{eq:deltam}
\end{equation}
When we apply this window, the priors are numerically renormalized.

\begin{figure}
\centering
\begin{subfigure}{0.45\textwidth}
    \includegraphics[width=\textwidth]{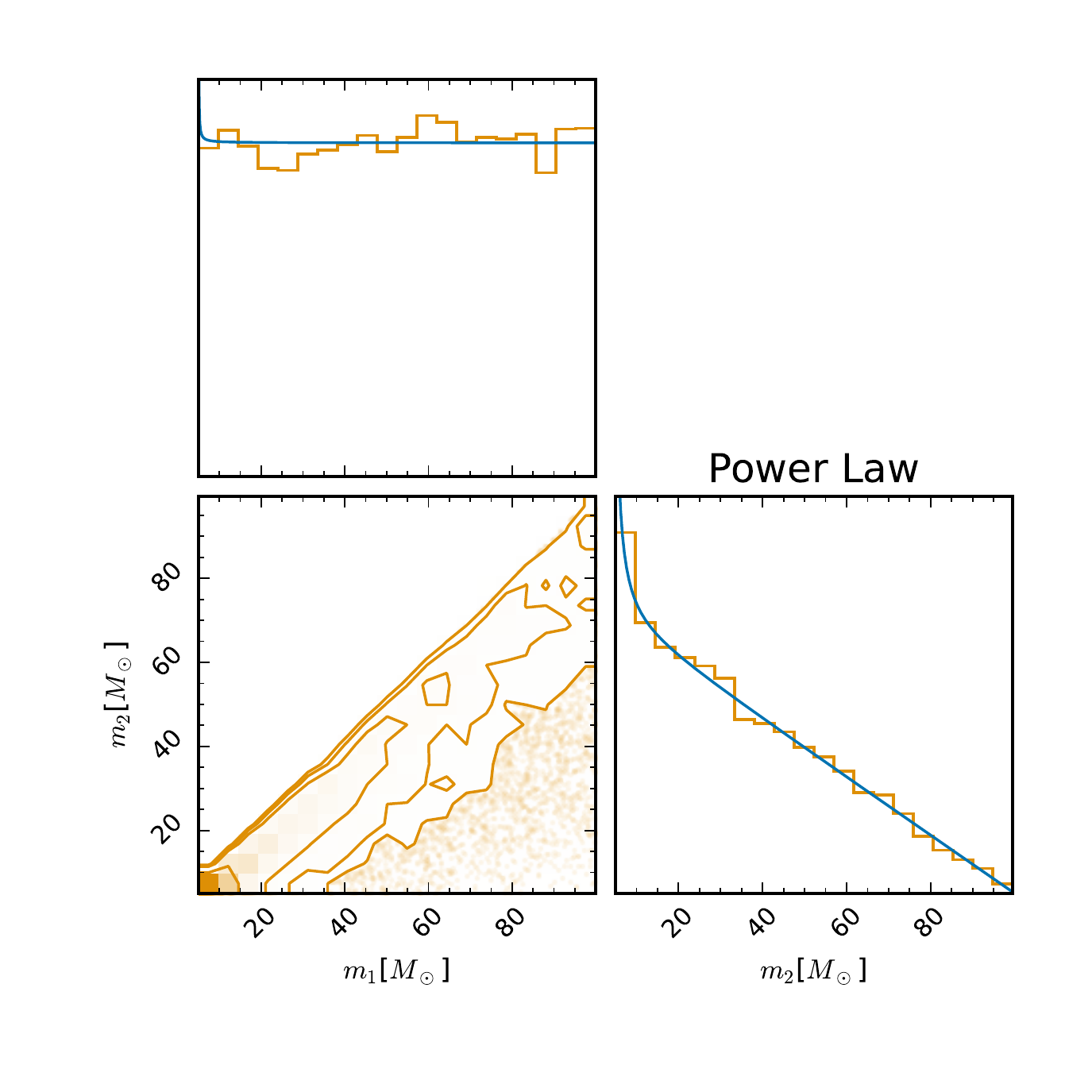}
    \caption{(a) Truncated Power Law model with parameters $\alpha=0$,$\beta=1$,$m_{\rm min}=5 M_\odot$,$m_{\rm max}=100 M_\odot$.}
    \label{fig:first}
\end{subfigure}
\hfill
\begin{subfigure}{0.45\textwidth}
    \includegraphics[width=\textwidth]{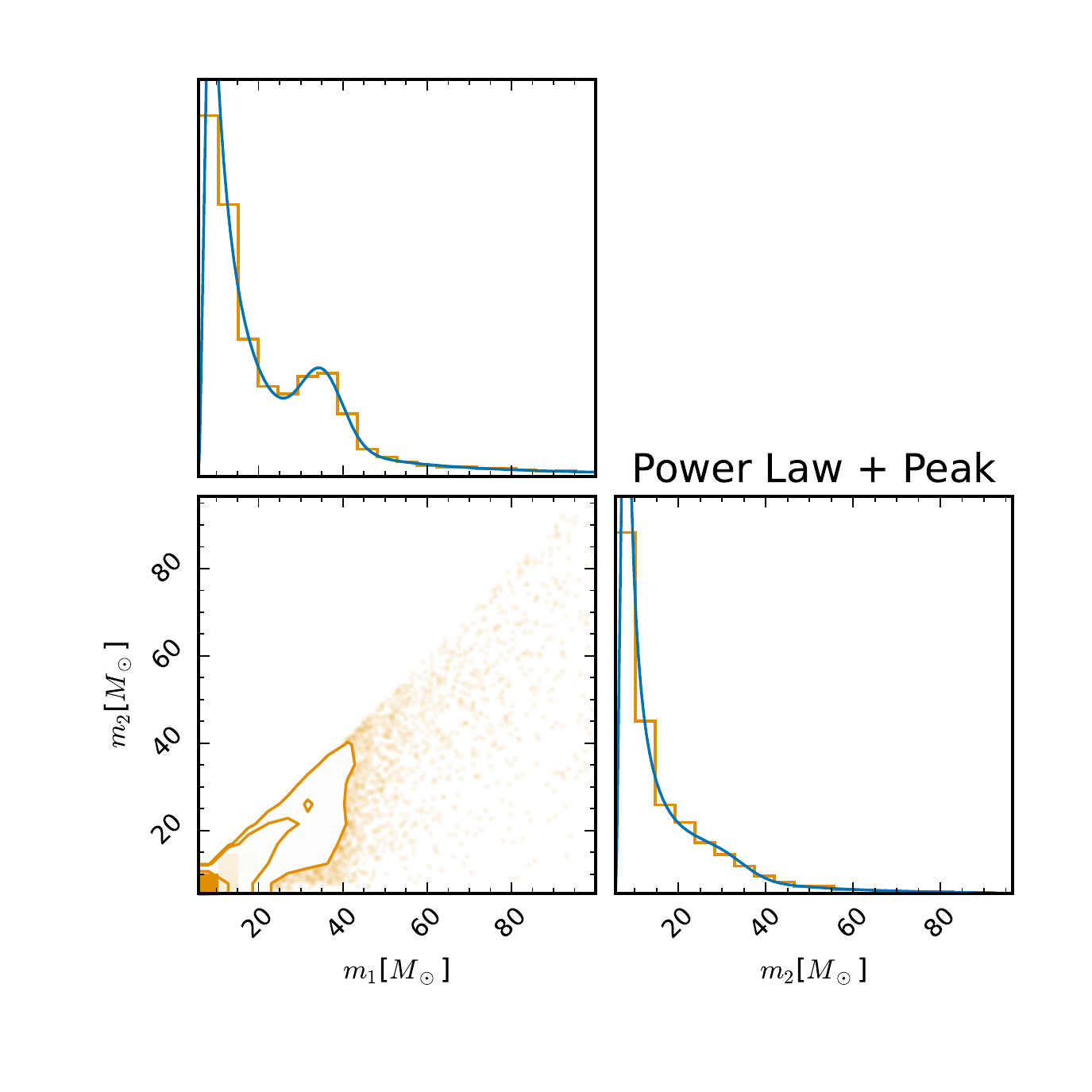}
    \caption{(b) \textsc{Power Law + peak} model with parameters $\alpha=2$,$\beta=1$,$m_{\rm min}=5 M_\odot$,$m_{\rm max}=100 M_\odot$,
          $\mu_g=35 M_\odot, \sigma_g=5 M_\odot, \lambda_p=0.1$}
    \label{fig:second}
\end{subfigure}
\hfill
\begin{subfigure}{0.45\textwidth}
    \includegraphics[width=\textwidth]{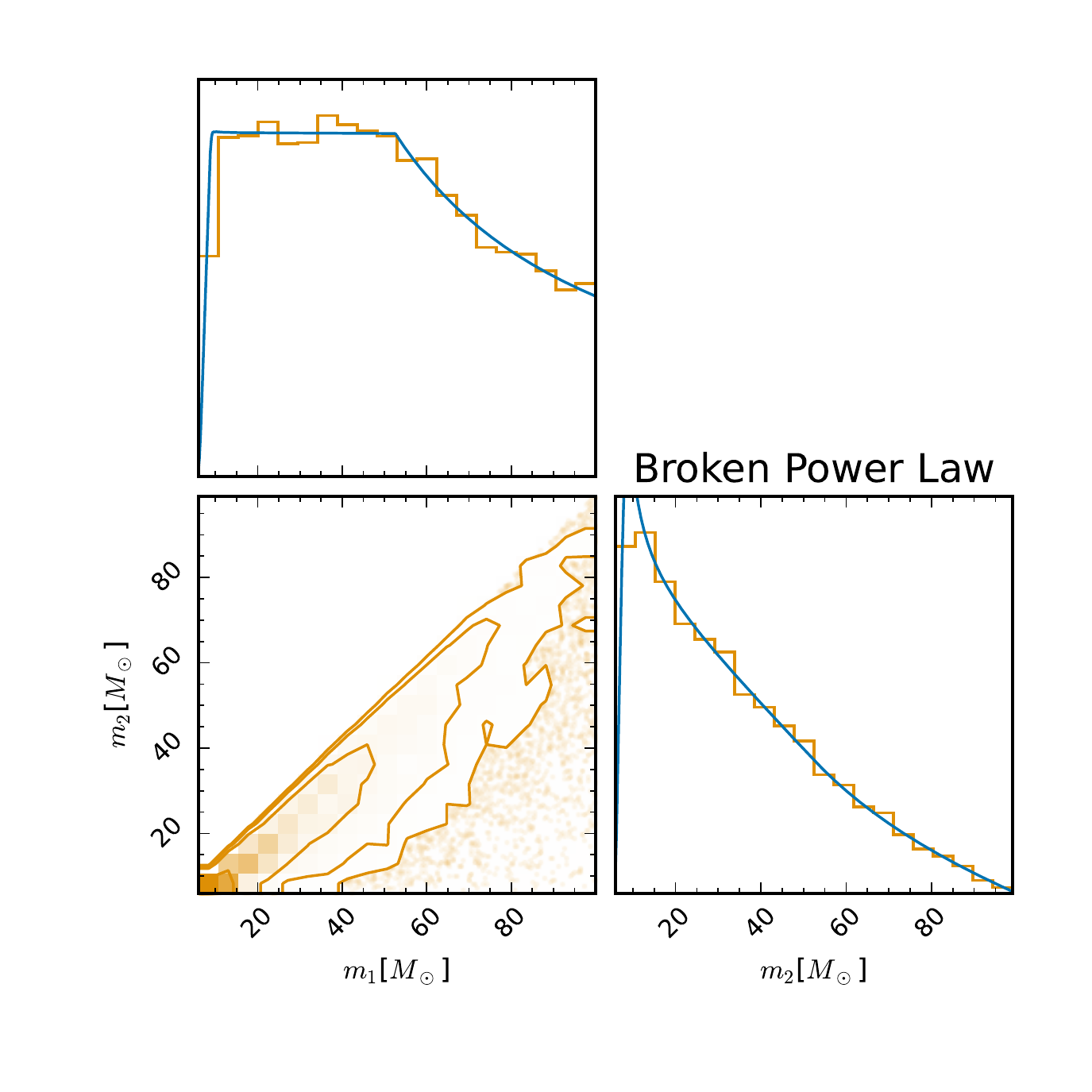}
    \caption{(c) \textsc{Broken power law} model with parameters $\beta=1$,$m_{\rm min}=5 M_\odot$,$m_{\rm max}=100 M_\odot$,$\alpha_1=0$,$\alpha_2=1$,$b=0.5$,$\delta_m=5 M_\odot$.}
    \label{fig:third}
\end{subfigure}
\hfill 
\begin{subfigure}{0.45\textwidth}
    \includegraphics[width=\textwidth]{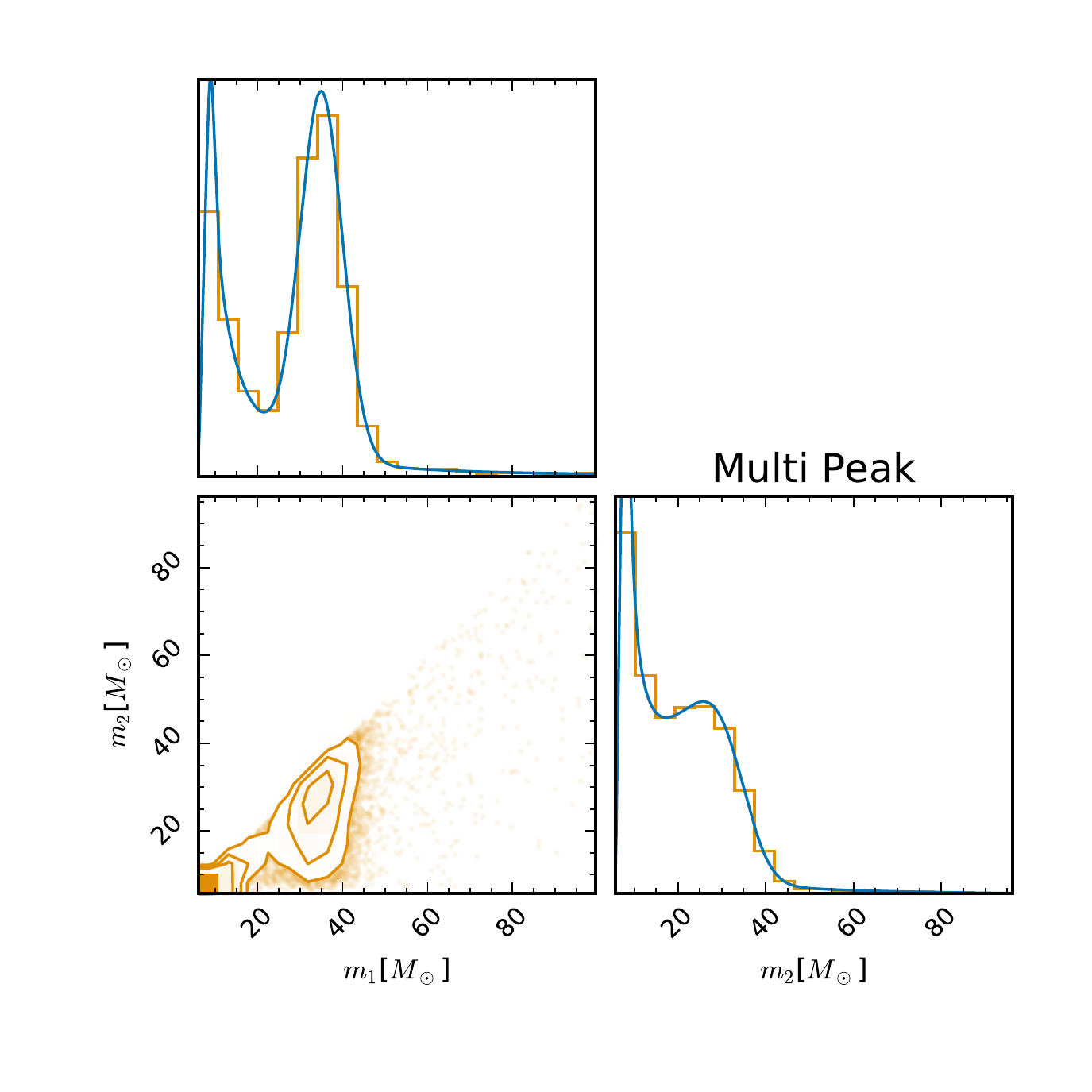}
    \caption{(d) \textsc{multi peak} model with parameters $\alpha=2,\beta=1,m_{\rm min}=5 M_\odot,m_{\rm max}=100 M_\odot,\delta_m=5 M_\odot,\mu_{\rm g,low}=9 M_\odot,\sigma_{\rm g,low}=1 M_\odot,\lambda_{\rm g,low}=0.05,\mu_{\rm g,high}=35 M_\odot,\sigma_{\rm g,high}=5 M_\odot,\lambda_g=0.5$}
    \label{fig:fourth}
\end{subfigure}

\caption{Sample of mass models implemented in \icarogw as distributions of primary and secondary source masses.}
\label{fig:mass models}
\end{figure}

All the mass models presented in this section can be visualized in Fig.~\ref{fig:mass models}.

\subsubsection{Truncated Power-Law}
The Truncated Power Law model is given by Eq.~(\ref{massprior0}) with:

\begin{eqnarray}
    \pi(m_{1,s}|m_{\rm min},m_{\rm max},\alpha)&=&\mathcal{P}(m_{1,s}|m_{\rm min},m_{\rm max},-\alpha)\,,\\
    \pi(m_{2,s}|m_{\rm min},m_{1,s},\beta)&=&\mathcal{P}(m_{2,s}|m_{\rm min},m_{1,s},\beta)\,,
\end{eqnarray}
where $\mathcal{P}$ is defined in Eq.~(\ref{TPL}).

\subsubsection{Power-Law + Peak}

This model was proposed in \citet{Talbot:2019okv} and it is given by Eq.~(\ref{massprior}) with:
\begin{eqnarray}
    \pi(m_{1,s}|m_{\rm min},m_{\rm max},\alpha)&=&(1-\lambda)\mathcal{P}(m_{1,s}|m_{\rm min},m_{\rm max},-\alpha)+\lambda \mathcal{G}(m_{1,s}|\mu_g,\sigma)\,, \quad (0 \leq \lambda\leq 1)\\
    \pi(m_{2,s}|m_{\rm min},m_{1,s},\beta)&=&\mathcal{P}(m_{2,s}|m_{\rm min},m_{1,s},\beta)\,.
\end{eqnarray}

\subsubsection{Broken Power Law}

This model is based on Eq.~(\ref{massprior}) and
consists basically of two truncated power-law distributions attached at the point $b$: 
\begin{equation}
    b = m_{\rm min} +(m_{\rm max}-m_{\rm min}) f,
\end{equation}
where  $f$ is a scalar in $[0,1]$, so $b=m_\mathrm{min}$ for $f=0$. This model was proposed in \citet{Abbott:2020gyp}.
 The priors are the following:
\begin{eqnarray}
    \pi(m_{1,s}|m_{\rm min},m_{\rm max},\alpha)&=&\frac{1}{N}[\mathcal{P}(m_{1,s}|m_{\rm min},b,-\alpha_1)+\frac{\mathcal{P}(b|m_{\rm min},b,-\alpha_1)}{\mathcal{P}(b|b,m_{\rm max},-\alpha_2))}\mathcal{P}(m_{1,s}|b,m_{\rm max},-\alpha_2)]\,,\\
    \pi(m_{2,s}|m_{\rm min},m_{1,s},\beta)&=&\mathcal{P}(m_{2,s}|m_{\rm min},m_{1,s},\beta)\,,
\end{eqnarray}
where the new normalization factor $N$ here is: 
\begin{equation}
    N=1+\frac{\mathcal{P}(b|m_{\rm min},b,-\alpha_1)}{\mathcal{P}(b|b,m_{\rm max},-\alpha_2))}.
\end{equation}

\subsubsection{Multi-Peak}
This model is based on Eq.~(\ref{massprior}) and consists of one power-law + two gaussian models 
with:
\begin{eqnarray}    
\pi(m_{1,s}|m_{\rm min},m_{\rm max},\alpha)&=&\Bigl[(1-\lambda)\mathcal{P}(m_{1,s}|m_{\rm min},m_{\rm max},-\alpha)+\lambda\lambda_{\rm low} \mathcal{G}(m_{1,s}|\mu_{g,\rm low},\sigma_{\rm low})+\\&&+ \lambda(1-\lambda_{\rm low}) \nonumber \mathcal{G}(m_{1,s}|\mu_{g,\rm high},\sigma_{\rm high})\Bigr]\,,
\\
 \pi(m_{2,s}|m_{\rm min},m_{1,s},\beta)&=&\mathcal{P}(m_{2,s}|m_{\rm min},m_{1,s},\beta)\,.
\end{eqnarray}
Note that since $\mathcal{G}$ and $\mathcal{P}$ are already normalized, $\pi(m_{1,s}|m_{\rm min},m_{\rm max},\alpha)$ is automatically normalised. The model was used in \citet{Abbott:2020gyp}.

\subsection{Spin models}
\label{app:spinsmod}

\begin{figure}[h!]
    \centering
    \includegraphics[scale=0.10]{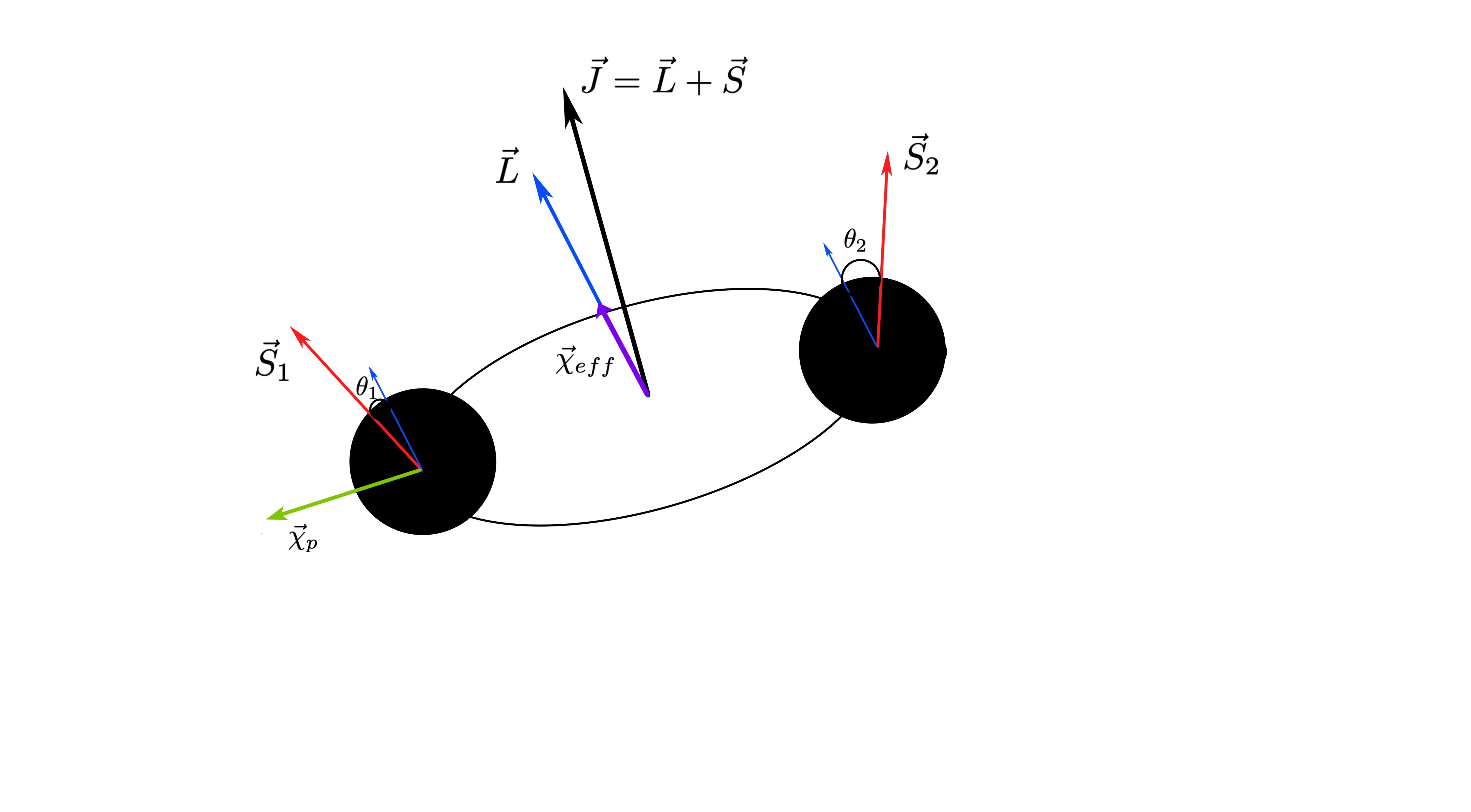}
    \vspace{-2cm}
    \caption{Representation of the spin components for a compact binary coalescence.}
    \label{fig:gaussian_model_scheme}
\end{figure}

In \icarogw we implemented two models for the CBC spins. The two models are based on two different parametrizations for the spin parameters of a binary. Referring to Fig.~\ref{fig:gaussian_model_scheme}, we provide a definition for the spin parameters typically employed in GW studies.

By definition, the $z$ axis of a binary is aligned to the instantaneous orbital angular momentum ${\vec{L}}$. 
The (normalized) spin amplitudes $\chi_{1,2}$, defined from the Cartesian components of the spin vectors, are:
\begin{eqnarray}
     \chi_1&=&\sqrt{s_{1,x}^2+s_{1,y}^2+s_{1,z}^2}\,,\\
     \chi_2&=&\sqrt{s_{2,x}^2+s_{2,y}^2+s_{2,z}^2}.
\end{eqnarray}
The tilt angles $\theta_{1,2}$ are defined as the angle between the BH spins and the orbital angular momentum, namely:
\begin{eqnarray}
     \cos \theta_1 &=& \frac{s_{1,z}}{\chi_1}\,, \\
     \cos \theta_2 &=& \frac{s_{2,z}}{\chi_2}. 
\end{eqnarray}
The \textit{effective spin parameter} $\chi_{\rm eff}$ and \textit{precession spin parameter} $\chi_{\rm p}$ are defined by (\cite{abbott2021population}):
\begin{eqnarray}
     \chi_{\rm eff} &=& \frac{\chi_1 \cos\theta_1 + q \chi_2 \cos\theta_2}{1+ q} = \frac{s_{1,z} + q s_{2,z} }{1+ q}\,, \\
     \chi_{\rm p} &=& {\rm max} \left[ \chi_1 \sin\theta_1; \,\left(\frac{4q +3}{3q +4}\right)q \chi_2 \sin\theta_2 \right],
\end{eqnarray}
where the mass ratio $q$ is:
\begin{equation}
    q=\frac{m_2}{m_1}\,, \qquad (q\leq 1).
\end{equation}

(The factors of $q$ appearing in the expression for $\chi_{\rm p}$ come from the leading order PN equation for $\dot{\vec{L}}$. Note that the 4 in-plane components of the perpendicular components of $\vec{s}$ have been replaced by one scalar $\chi_{\rm p}$ which is an averaged quantity, see Eq.~(3.1) of~\cite{Schmidt:2014iyl}).

\begin{figure}
\centering
\begin{subfigure}{0.45\textwidth}
    \includegraphics[width=\textwidth]{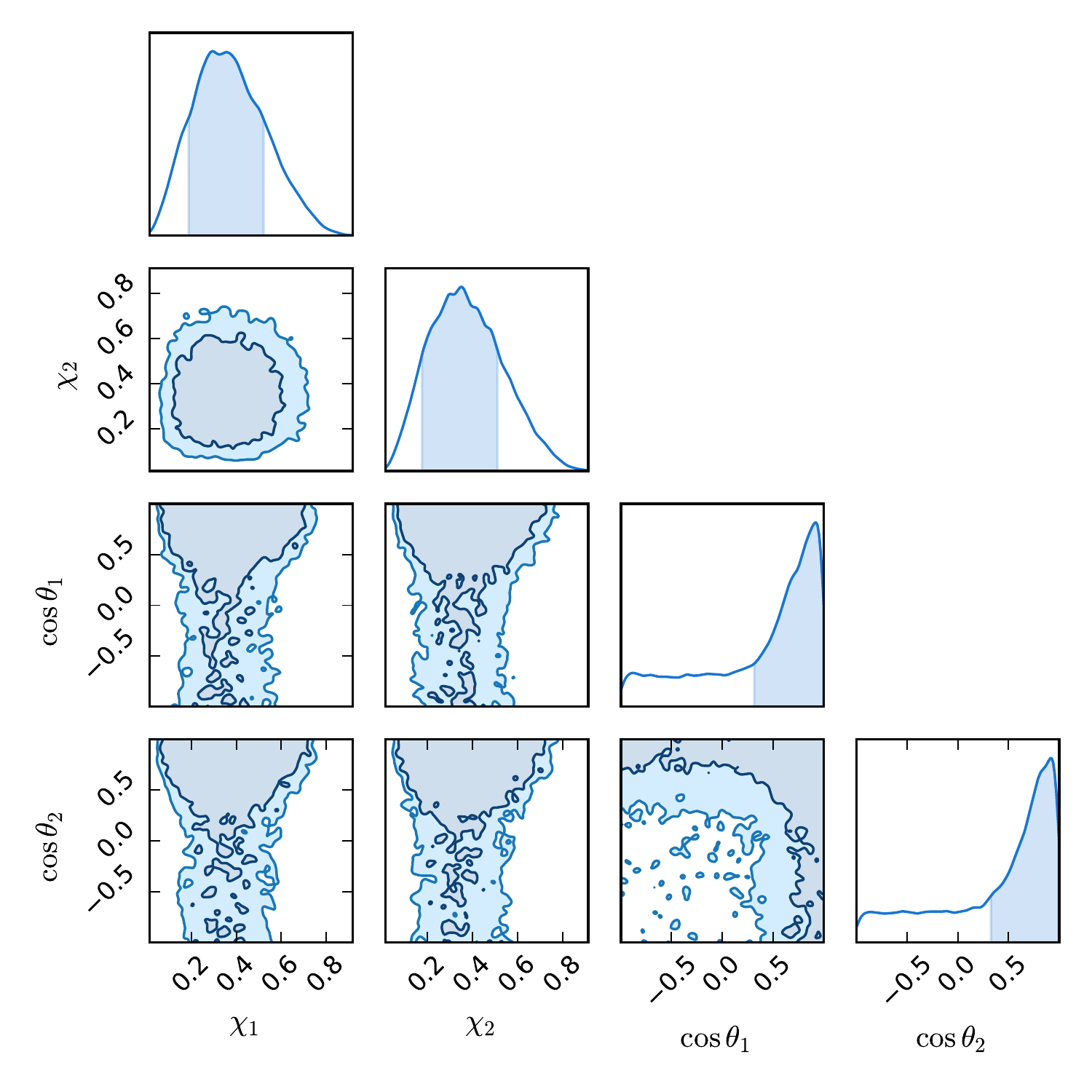}
    \caption{(a) \textsc{default} spins model with $ \alpha_\chi=3,\beta_\chi=5,\sigma_t=0.3,\xi=0.5$}
    \label{fig:first}
\end{subfigure}
\hfill
\begin{subfigure}{0.45\textwidth}
    \includegraphics[width=\textwidth]{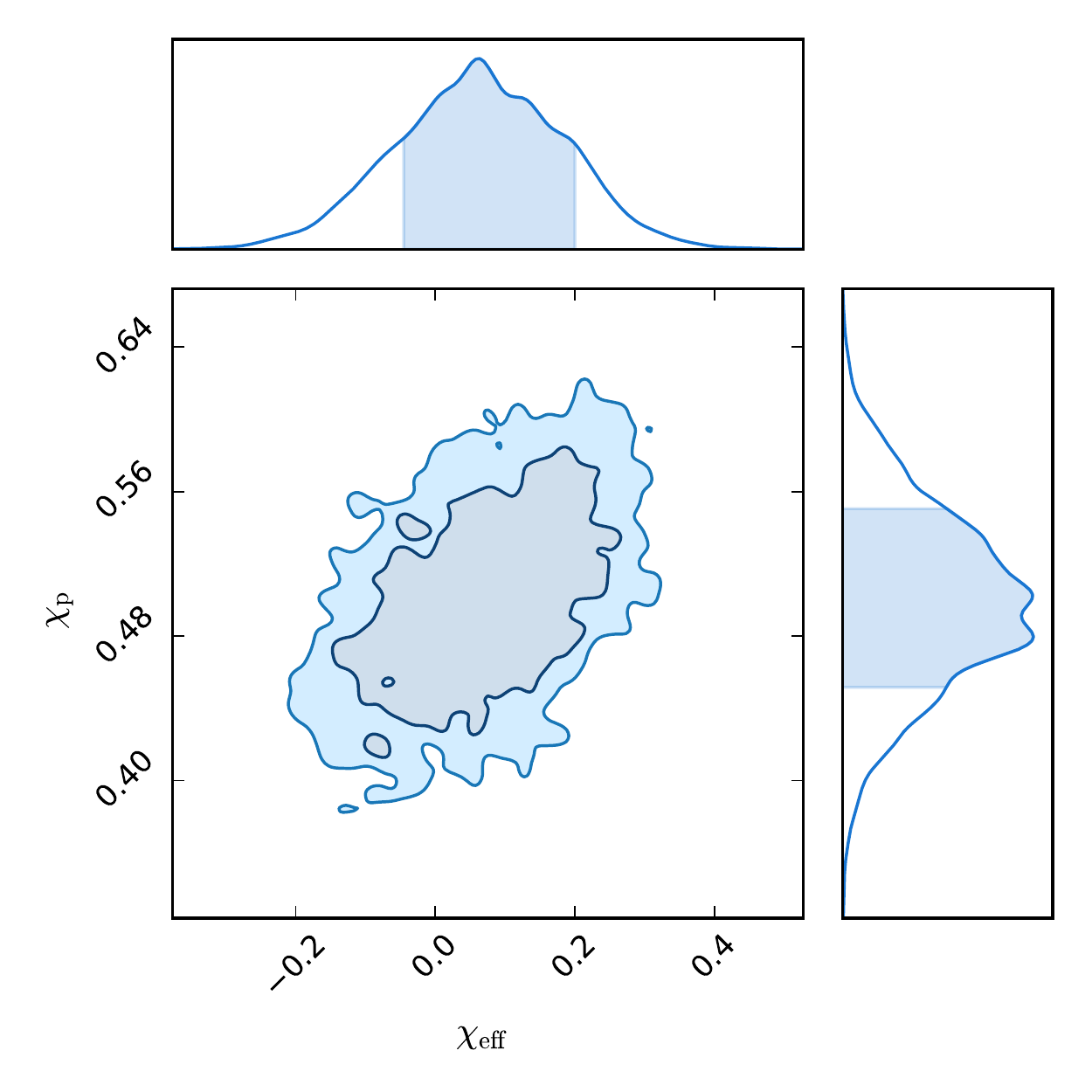}
    \caption{(b) \textsc{gaussian} spins model with $\mu_{\chi_{\rm eff}}=0.06,\sigma_{\chi_{\rm eff}}=0.12,\mu_{\chi_{\rm p}}=0.5,\sigma_{\chi_{\rm p}}=0.05,\rho=0.5$.}
    \label{fig:second}
\end{subfigure}
\caption{Sample of spin models implemented in \icarogw as a function of different spin parameters.}
\label{fig:spins}
\end{figure}

The parameter $\chi_{\rm eff}$ accounts for the amount of spin aligned with the orbital angular momentum, as well as the magnitude of the BH spins. Since $\chi_{\rm eff}$ is bounded between $[-1,1]$, values close to $\chi_{\rm eff}= 1$ correspond to highly spinning BHs with aligned spins, whereas values close to $\chi_{\rm eff}=-1$ support highly spinning BHs with anti-aligned spins, and $\chi_{\rm eff}=0$ is consistent with non spinning BHs. The \textit{precession spin parameter} $\chi_{\rm p}$, bounded between $[0,1]$, quantifies the amount of spin perpendicular to the angular momentum.  All the spin models described below can be visualized in Fig.~\ref{fig:spins}.

\subsubsection{\textsc{Default} spin model}
The model was used in \citet{Abbott:2020gyp} and it is proposed after \citet{Talbot:2019okv,PhysRevD.100.043012}. The model works with spins parameterized using the two spin magnitudes $\chi_1,\chi_2$ and the two cosine of inclination angles 
$\cos t_1, \cos t_2$ with respect to the orbital angular momentum. The total number of degrees of freedom (d.o.f) of a BBH system in terms of spin is 6. The last two remaining d.o.f are the azimuthal angles $\phi _1$ and $\phi _2$ are not considered here and supposed uniform. 

The population distribution is given by: 
\begin{equation}
    \pi(\chi_1,\chi_2,\cos \theta_1,\cos \theta_2) = {\rm Beta}(\chi_1|\alpha,\beta) \pi(\cos \theta_1|\xi,\sigma_t) {\rm Beta}(\chi_2|\alpha,\beta) \pi(\cos \theta_2|\xi,\sigma_t) ,
\end{equation}
namely it is factored into two parts. Above ``Beta'' is the beta distribution, calculated with parameters $\alpha$ and $\beta$ defined by:
\begin{eqnarray}
     \alpha &=& \left(\frac{1-\mu_\chi}{\sigma^2_\chi} -\frac{1}{\mu_\chi}\right)\mu_\chi^2 \geq 1\,, \label{eq:B45}\\
    \beta &=& \alpha\left(\frac{1}{\mu_\chi} -1\right) \geq 1 \label{eq:B46}.
\end{eqnarray}

The condition $(\alpha,\beta)\geq 1$ is imposed to avoid any non-singular asymptotic behavior of the Beta distribution. The probability density function for the angle distribution is given by (see Eq. (14) in \cite{abbott2021population}): 
\begin{equation}
    \pi(\cos \theta_{1,2}|\zeta,\sigma_t)=\xi \mathcal{G}_{[-1,1]}(\cos \theta_{1,2}|1,\sigma_t) +\frac{1-\xi}{2},
    \label{eq:B47}
\end{equation}
where $\mathcal{G}_{[-1,1]}(\cos \theta_i|1,\sigma_t)$ (see Eq.\eqref{eqn:trGauss}) is a  truncated gaussian between $-1$ and $1$ on $\cos \theta_i$, with mean $1$ and standard deviation $\sigma_t$. \newline
\textbf{Note:} the parameter $\xi$ is bounded between $[0,1]$: the form of the angle distribution is a mixed model between a truncated gaussian and a uniform distribution between $-1$ and $1$, where $\xi$ is the mixing parameter.

\subsubsection{\textsc{Gaussian} spin model}

The \textsc{Gaussian} spin model seeks to measure the joint distribution of $\chi_{\rm eff}$ and $\chi_p$. It was proposed in \citet{Miller_2020} and it depends on 5 parameters that are $\mu_{\chi_{\rm eff}}, \sigma_{\chi_{\rm eff}}, \mu_{\chi_{\rm p}}, \sigma_{\chi_{\rm p}}$, and $\rho$.  The population probability on $\chi_{\rm eff},\chi_{\rm p}$ is a bivariate gaussian truncated between [-1,1] for $\chi_{\rm eff}$ and between $[0,1]$ for $\chi_{\rm p}$. The covariance of the bivariate gaussian is ${\rm cov}_{[\chi_{\rm eff},\chi_{\rm p}]}= \rho \sigma_{\chi_{\rm eff}} \sigma_{\chi_{\rm p}}$.
In \icarogw, this bivariate gaussian is factorized as:
\begin{equation}
    \pi(\chi_{\rm eff},\chi_{\rm p}|\mu_{\chi_{\rm eff}}, \sigma_{\chi_{\rm eff}}, \mu_{\chi_{\rm p}}, \sigma_{\chi_{\rm p}},\rho)=\mathcal{G}_{[-1,1]}(\chi_{\rm eff}|\mu_{\chi_{\rm eff}}, \sigma_{\chi_{\rm eff}}) \mathcal{G}_{[0,1]}(\chi_{\rm p}|\mu_*,\sigma_*),
    \label{eq:gmodel}
\end{equation}
where 
\begin{eqnarray}
   \mu_*&=&\mu_{\chi_{\rm p}}+\frac{{\rm cov}_{[\chi_{\rm eff},\chi_{\rm p}]}}{\sigma^{2}_{\chi_{\rm eff}}} (\chi_{\rm eff}-\mu_{\chi_{\rm eff}})\,, \\
   \sigma_*&=& \frac{\sigma_{\chi_{\rm p}} {\rm cov}_{[\chi_{\rm eff},\chi_{\rm p}]}}{\sigma^{2}_{\chi_{\rm eff}}}\,.
\end{eqnarray}

This factorization is equivalent to a bivariate gaussian distribution.

\bibliographystyle{aa} % style aa.bst
\bibliography{bib} % your references Yourfile.bib

\end{document}